\documentclass[aps,pre,twocolumn,superscriptaddress,showpacs]{revtex4}
\usepackage{graphicx,amssymb,amsmath,mathrsfs,color}

\begin{document}

\title{Anomalous diffusion in time-fluctuating non-stationary diffusivity
landscapes}

\author{Andrey G. Cherstvy}
\email{a.cherstvy@gmail.com}
\affiliation{Institute for Physics \& Astronomy, University of Potsdam, 14476
Potsdam-Golm, Germany}

\author{Ralf Metzler}
\email{rmetzler@uni-potsdam.de}
\affiliation{Institute for Physics \& Astronomy, University of Potsdam, 14476
Potsdam-Golm, Germany}

\begin{abstract}

We investigate the ensemble and time averaged mean squared displacements for particle diffusion in a simple model for disordered media by assuming that the local diffusivity is both fluctuating in time and has a deterministic average growth or decay in time. In this study we compare computer simulations of the stochastic Langevin equation for this random diffusion process with analytical results. We explore the regimes of normal Brownian motion as well as anomalous diffusion in the sub- and superdiffusive regimes. We also
consider effects of the inertial term on the particle motion. The investigation of
the resulting diffusion is performed for unconfined and confined motion. 

\end{abstract}

\pacs{05.40.a}

\date{\today}

\maketitle

\section{Introduction}

The diffusion of a tracer particle is typically characterised in terms of the
mean squared displacement (MSD) \begin{equation} 
\left< x^2(t)\right>=\int x^2P(x,t)dx=2K_{\gamma}t^\gamma
\label{eq-ad} \end{equation} corresponding to the second moment 
of the probability density function $P(x,t)$
to find the particle at position $x$ at time $t$. When the exponent $\gamma=1$
the law (\ref{eq-ad}) describes normal Brownian diffusion, otherwise we speak
of anomalous diffusion. In the latter case, the generalised diffusion coefficient
$K_{\gamma}$ has the physical dimensions $\mathrm{cm}^2/\mathrm{sec}^{\gamma}$,
and we distinguish subdiffusion ($0<\gamma<1$) and superdiffusion ($\gamma>1$)
depending on the value of the anomalous diffusion exponent $\gamma$ \cite{metz00}.

Following a surge in microscopic techniques, diffusive phenomena of passive tracer
particles can now be monitored at unprecedented resolution \cite{brauchle}. Thus,
for instance, the hydrodynamic backflow effects of a Brownian particle could be
directly probed \cite{jeney}. Even more remarkable is the rapidly growing number
of experimental evidence for anomalous diffusion in dense fluids \cite{dense} as
well as in living biological cells \cite{metz11,metz12,fran13, metz14,soko15,para15}.
The motion of various endogenous and artificial tracers in live cells was shown
to be subdiffusive \cite{cell,weis04,metz11b}. However, when active dynamics such as
driving by molecular motors or cytoplasmic streaming are involved, superdiffusion
may also be observed \cite{cellsuper}. Massive computer simulations of pure and
protein-crowded lipid bilayer membranes demonstrate transient anomalous diffusion
of both lipids and proteins, the crossover to normal diffusion being delayed with
increasing disorder \cite{lipids,prx}.  In the membranes of living cells anomalous
diffusion is observed on macroscopic time scales \cite{membranes,lape15,metz16}.
As a general physical principle for anomalous diffusion various form of crowding of
the environment are considered \cite{crowd,ghos15,ghos16}. We note that anomalous diffusion also
occurs on the level of entire organisms, such as the subdiffusion of bacteria cells
in biofilms \cite{rogers} or the superdiffusion of hydra or protozoa \cite{hydra}.

Several additional studies of crowded \textit{in vitro} systems demonstrate the existence
of non-Fickian and/or non-Gaussian motion, for instance the glassy dynamics in
membrane domains \cite{mung16}, confined diffusion of water molecules in soft
environments \cite{moss16}, polymer diffusion on nanopillar-structured surfaces
\cite{schw15}, intermittent molecular hopping on solid-liquid interfaces
\cite{schw13}, colloidal spheres in dense crowded suspensions \cite{yeth14} and
glasses \cite{week00, zamp14}, particle diffusion in porous media with
heterogeneous and position-dependent mobilities \cite{schw15b}, and the transport
of contaminants in porous and fractured geological formations \cite{berk06}.
Concurrently the existence of \emph{anomalous yet Brownian} diffusion---a linear
time dependence of the MSD (\ref{eq-ad}) accompanied by
significantly non-Gaussian (exponential or stretched exponential) probability
density $P(x,t)$---was demonstrated for the motion of colloidal beads along linear
phospholipid bilayer tubes \cite{gran09,gran12}, particle dynamics in hard
sphere colloidal suspensions \cite{gran14}, and diffusion of nanoparticles
in nanopost arrays \cite{kris13}.

Brownian motion is bound to the Gaussian shape of the probability density function
by the spell of the central limit theorem and thus fully characterised by the
second moment (\ref{eq-ad}). In contrast, anomalous diffusion dynamics is inherently
non-universal, and therefore a large variety of {anomalous diffusion models exists (also with non-Gaussian probability densities), depending on the exact physical circumstances defining the dynamics \cite{metz00, metz11,fran13,metz14, soko15,para15}. To name but a few of these anomalous diffusion processes we recall continuous time random walks with scale free trapping time distributions \cite{mont65} and a potential additional noise source \cite{noisy}, general trapping models \cite{traps}, correlated diffusion processes \cite{corr,
corr1}, fractional Brownian motion \cite{fbm} and generalised Langevin equation
motion \cite{fbm,fle,pagn16,bark09}, as well as diffusion in disordered media and
on fractal structures \cite{kehr87,bray13,mard15}.

Here we focus on models based on randomly and/or deterministically varying diffusion
coefficients which have recently been under intense study. We show that when the
diffusion coefficient varies randomly such that its distribution has a finite
width, normal diffusion emerges in the long time limit. We analyse these random diffusion processes (RDPs) in terms of the MSD and time averaged MSD typically evaluated in single particle tracking and simulations studies, for both situations of unconfined and confined diffusion. In addition, we quantify the degree of randomness between different individual trajectories.

In section \ref{hdp} we provide a concise overview of heterogeneous diffusion
processes, followed by a definition of the various observables for (anomalous)
diffusion processes in section \ref{obs}. In section \ref{sec-model} we present
the details of the specific model investigated here and the numerical scheme used
to simulate the RDPs. We then present the main results of our calculations for
the ensemble averaged MSD, time averaged MSD, probability distribution function
$P(x,t)$, and the ergodic behaviour of RDPs in sections \ref{sec-results-over} and
\ref{sec-results-under}, respectively, for massless and massive particles. We
consider the situation both in the absence and in the presence of an external
confinement. Section \ref{sec-disc} summaries our findings and discusses their
possible applications and generalisations.

\section{Heterogeneous mobility: models and examples}
\label{hdp}

\subsection{Random diffusivity models}

For massless particles the study of normal and anomalous diffusion in the
presence of random diffusivity fields recently attracted considerable
attention \cite{slat14,lape14, lape15,akim15,seba16}. Several models assume
that the instantaneous diffusion coefficient $D$ is governed by a steady
state distribution $p(D)$ in an annealed fashion, that is, the instantaneous of
$D$ is independent of the actual particle position $x$. Particular attention
received the idea of a \emph{diffusing diffusivity} introduced by Chubynsky
and Slater \cite{slat14}. In their model they assume an exponential distribution
\begin{equation}p(D)=D_0^{-1}\exp[-D/D_0]\label{eq-expon-chuby}\end{equation} 
and weight the standard Gaussian $P(x,D,t)=(4\pi Dt)^{-1/2}\exp(-x^2/[4Dt])$ with this function, $\mathfrak{P}(x,t)=\int_0^{\infty}p(D) P(x,D,t)dD,$ to obtain the exponential probability distribution function (PDF) \begin{equation} \mathfrak{P}(x,t)=\frac{1}{\sqrt{4D_0t}}\exp\left[-\frac{|x|}{\sqrt{D_0t}}\right]. \end{equation} The characteristic decay length of this PDF grows with the diffusion time as $\simeq t^{1/2}$ \cite{slat14}, but the MSD grows linearly with time, that is, follows Eq.~(\ref{eq-ad}) with $\gamma=1$ and $K_{\gamma}=D_0$ \cite{slat14}. Other than exponential diffusivity distributions---for instance, power-law forms of $p(D)$---were shown to lead to subdiffusive and non-ergodic MSD behaviour \cite{lape14}. After the current manuscript was submitted, the authors became aware of the simulation-based study \cite{egel16} of particle 
diffusion in rough energy landscapes with both Gaussian and Gamma distributed local energy values.

Dynamical heterogeneities---as reflected in the above assumption of a random
diffusivity---are considered a characteristic property of systems such as
supercooled or glassy liquids \cite{akim15,yama98, rich02,gotz92,biro11}.
Quite broad distributions of particle diffusivities were detected in a number
of living systems, for instance, for the motion of pathogen receptors on two
dimensional cell membranes \cite{lape15} (see also Ref. \cite{saxt97}), 
motion of Cajal bodies in eukaryotic nuclei \cite{platani}, one dimensional diffusion of repressor proteins on the DNA \cite{aust06}, and for the motion of proteins along the
corrugated landscape created by the DNA sequence \cite{goyc14,baue15}. In these
systems, the \textit{inherent stochasticity} of the diffusive properties of a tracer
particle as well as the heterogeneities of its environment contribute to the
observed distribution of diffusivities subsumed in the distribution $p(D)$. 

The reader is particularly referred to the characterisation of the dynamical
spreading of a population of nematode worms, both in homogeneous and
heterogeneous environments \cite{hapc09}. Other examples of living systems
with diffusing individuals obeying non-Gaussian distribution of diffusivities,
speeds of motion $v$, or turning angles are also mentioned in this study. Hapca
et al. \cite{hapc09} state that the anomalous diffusion monitored in the
heterogeneous populations of worms can be solely due to fat-tailed, e.g.,
gamma distributed \cite{yama02} forms of diffusivities, while the motion of
each individual remains Brownian. This study served as a strong biological
motivation for us in trying to unveil the properties of particle diffusion
with a given time dependent form of $p(D,t)$.

Also note that fat-tailed leptokurtic distribution of particle mobilities
often occurring in population of individuals of a species can ensure a
facilitation of their colonial invasion \cite{hapc09}, as compared to the
standard Brownian diffusion law of spreading. Such skewed distributions $p(D)$
or $p(v)$ (of the particle speeds) can originate from medium heterogeneities
when the organisms explore different regions of space with different mobilities.
The notion of fat-tailed distributions and faster than standard front propagation
emerges also in long-distance dispersal of plants \cite{clar98,nath06} and pollen
\cite{yama04}, in patterns of fish movements \cite{skal00}, as well as in
rare event driven spreading of plant pathogens \cite{hovm02}.

\subsection{Deterministic variation of the diffusivity with position or time}

Following experimental observations of deterministic gradients of the local
diffusivity in both pro- and eukaryotic cells \cite{lang11,elf11} and
the existence of thermal gradient conditions \cite{brau13}, the model of
heterogeneous diffusion processes (HDPs) with a power-law, exponential, and
logarithmic form for $D(x)$  was recently introduced by the authors \cite{cher13,cher14a,cher14b}; see also Refs. \cite{lenz03,fa05,zimm11,heid14}. These Markovian processes based on a Langevin description with multiplicative noise exhibit anomalous diffusion and weak ergodicity breaking \cite{cher13,cher14a,cher14b}. The latter emerges due to the fact that even in the limit of long trajectories time and ensemble averages of physical observables do not coincide \cite{metz12,metz14}, see
below. Models with a power-law time dependence \begin{equation} D(t)\simeq t^{\alpha-1} \label{eq-DofT} \end{equation} of the diffusivity exist, 
the so-called scaled Brownian motion (SBM) \cite{muni02,
pein11,soko14,jeon14,safd15,grig16, metz14,bodr15gg}. SBM was originally introduced
by Batchelor in the description of Richardson turbulence \cite{batc52}. Note that
in the context of such highly non-stationary processes the degree of ergodicity
breaking is controlled via introducing a time- or length-dependent scale into the
problem \cite{pagn16,cher15c}. A combined space-time diffusivity dependence of the form $D(x,t)\simeq|x|^{\beta}t^{\alpha-1}$ was also investigated \cite {fuli13,fuli11,cher15b}. Aged and confined
versions of these processes were recently considered as well \cite {cher14b,jeon15}. 
Here ageing refers to the explicit
dependence of the process on the overall time span of its evolution, as expected for non-stationary processes.
The limiting cases with scaling exponent $\beta\to2$ for HDPs and $\alpha\to0$
for SBM were shown to lead, respectively, to an ultrafast (exponential) and
ultraslow (logarithmic) MSD growth with time \cite{cher14c,lube07,bodr15}.

\subsection{Massless versus Massive Particles}

For massive particles, the situation in anomalous diffusion field is often less
clear. While the underdamped limit of standard Brownian motion \cite{uhle30} and
of fractional Langevin equation motion \cite{bark09} are well understood, for
other anomalous diffusion processes these limits and general solutions are just
emerging. In particular, for SBM it was recently shown that the long time limit
of underdamped motion (including the inertia term) does not always correspond to
the overdamped limit of the same motion \cite{bodr16}. Also, the recent study \cite{soko16} addresses a giant particle diffusion in the underdamped limit with a temperature dependent diffusion coefficient and in the presence of a bias.
One purpose of the current study is to investigate RDPs for massive and massless particles.
A growing interest in the diffusive behaviour of tracked particles combined with
the unprecedented precision of experimental observations, in particular, at short
times for the diffusion of small particles in living cells \cite{rien14}, pose a
need for the development of new and more flexible models of stochastic processes.
Thus, a larger pool of theoretical models is necessary for quantitative
descriptions of these systems, with possibly fewer number of model parameters.

Some implications of a finite particle mass for diffusion processes with position
dependent diffusivity of the form $D(x)\simeq|x|^\beta$ were recently examined
\cite{fara17}. The reader is also referred to the studies \cite{fara14,fara14b}
regarding the inertial Langevin dynamics in media with space inhomogeneous
friction, and conventions of how to interpret the associated multiplicative
stochastic equation as well as the existence of fluctuation-dissipation relations
for such systems. In what follows, we refer to the diffusion coefficient $D$ as to
local variable in space and time, rather as to a long time asymptote of the
Einstein relation, see the discussion in Ref. \cite{fara14b}.

\section{Observables of diffusion processes}
\label{obs}

Anomalous diffusion processes can be classified by the MSD diffusion 
exponent $\gamma$. In single particle tracking and simulations studies garnering few but long individual
time series $x(t)$ of the particle position the time averaged MSD \cite{metz12,
metz14}
\begin{equation}
\label{eq-tamsd}
\overline{\delta^2(\Delta)}=\frac{1}{T-\Delta}\int_0^{T-\Delta}\Big[x(t+\Delta)
-x(t)\Big]^2 dt
\end{equation}
is however typically employed. Here $T$ is the total length of the trajectory (observation
time) and $\Delta$ is the lag time. Note that while the ensemble averaged MSD
(\ref{eq-ad}) is a spatial average at a particular time instant $t$, the time
averaged MSD (\ref{eq-tamsd}) for any given lag time $\Delta$ is taken over the
entire history of the trajectory $x(t)$. As usual, ensemble averaging is denoted
hereafter by angular brackets, while time averaging is indicated by the overline.
To obtain smoother curves for the time averaged MSD an additional average is 
taken over $N$ trajectories, defining the mean time averaged MSD \cite{metz12,metz14}
\begin{equation}
\left<\overline{\delta^2(\Delta)}\right>=\frac{1}{N}\sum_{i=1}^N\overline{\delta^2
_i(\Delta)}.
\label{eq-mean-tamsd}
\end{equation}

Ergodicity in the Boltzmann-Khinchin sense typically assumed in equilibrium
statistical mechanics would imply the equivalence of ensemble and time averaged
MSD in the limit of long measurement times, $\lim_{T\to\infty}\overline{\delta
^2(\Delta)}=\langle x^2(\Delta)$. Following Bouchaud \cite{bouchaud_web} the
breakdown of this relation is referred to as weak ergodicity breaking
\cite{metz08,soko08,buro10,metz11,metz14,weak},
\begin{equation}
\lim_{T\to\infty}\overline{\delta^2(\Delta)}\neq\left<x^2(\Delta)\right>.
\label{eq-nonequiv}
\end{equation}
Continuous time random walks and HDPs are known to be weakly non-ergodic
\cite{metz14}, while diffusion on fractals is ergodic on the infinite cluster but
not on the entirety of all clusters \cite{mard15}. In contrast, other diffusive
processes such as fractional Brownian motion and SBM are only marginally
non-ergodic \cite{jeon10,metz14,soko14,jeon13,safd15,gode16,kurs13,greb12,jeon14}.

A distinctive measure of non-reproducibility of individual time averaged MSD traces 
is the ergodicity breaking parameter \cite{metz08}
\begin{equation}
\mathrm{EB}(\Delta)=\left<\xi^2(\Delta)\right>-1
\label{eq-eb-general}
\end{equation} 
based on the dimensionless ratio $\xi(\Delta)=\overline{\delta^2(\Delta)}/\left<
\overline{\delta^2(\Delta)}\right>$ quantifying the spread of individual time
averaged MSDs about their mean (\ref{eq-mean-tamsd}). Typically EB of a weakly
non-ergodic process decays to zero with increasing trace length slower 
than for the standard Brownian motion \cite{bark09,greb12},
\begin{equation}
\lim_{T\to\infty}\mathrm{EB}_{\mathrm{BM}}(\Delta)=\frac{4\Delta}{3T}.
\label{eq-eb-bm}
\end{equation}
Or, EB may even attain a finite value as $\Delta/T\to0$, for instance, for HDPs
and continuous times random walks \cite{metz08,soko08,cher13,metz14}. Often, also
the ratio of the time and ensemble averaged MSDs \cite{gode13}
\begin{equation}
\mathcal{EB}(\Delta)={\left<\overline{\delta^2(\Delta)}\right>}/{\left<x^2(
\Delta)\right>}
\label{eq-eb2}
\end{equation}
provides additional information about the ergodic properties of the diffusion
process.

\section{The random diffusivity model}
\label{sec-model}

In this section we describe the details of RDPs. As a generalisation of Eq. (\ref{eq-expon-chuby}), the instantaneous value of the diffusion coefficient on each simulation step is
independently chosen from the Rayleigh distribution
\begin{equation}
p(D)=\frac{D}{D_\sigma^2}\exp\left[-\frac{D^2}{2D_\sigma^2}\right].
\label{eq-pD-Rayleigh}
\end{equation}
The mean particle diffusivity is then given by
\begin{equation}
\left<D\right>=\sqrt{\pi/2} \times D_\sigma
\label{eq-average-D}
\end{equation}
and the diffusivity variance is $\left<(D-\left<D\right>)^2\right>=(2-\pi/2)D_\sigma^2$. Independence of successive
values of the diffusion coefficient indicates \textit{no temporal correlations} in its fluctuations. Our system
is thus out of equilibrium (the temperature is not fixed) and the fluctuation
dissipation theorem does not hold. The PDF (\ref{eq-pD-Rayleigh}) is a smooth function in the range from  $D=0$ to $D=\infty$ and it vanishes on the boundaries of this interval. This $p(D)$ distribution is used instead of a Gaussian distributed diffusion coefficient to avoid non-physical negative $D$ values. The distribution (\ref{eq-pD-Rayleigh}) is thus physically different from the exponential $p(D)$ form utilised of Chubynsky and Slater in Ref. \cite {slat14}. When the mean diffusivity stays constant over time, in the long time limit
the particle diffuses normally. However, when the mean diffusivity varies as a power-law, 
\begin{equation} \left<D(t)\right> \simeq t^\omega,\label{eq-sigma-power-law}\end{equation}
the resulting process is reminiscent of SBM. Physically, such an increase of the mean diffusivity could be due to a \emph{diffusing diffusivity\/} of the form $\langle D(t_i)\rangle
=|\langle D(t_{i-1})\rangle+\zeta_D(t_{i-1})|$, where $\zeta_D(t_i)$ is an
incremental change. This is analogous to the power-law growth of the waiting times in
the correlated continuous time random walk \cite{corr1}.

We consider below both massless and massive particles diffusing in both 
unconfined and a confined environments. We implement the same algorithms for
the iterative computation of the particle displacement $x(t)$ as developed for
HDPs \cite{cher13} and combined HDP-SBM motion \cite{cher15b}. First, we simulate the one dimensional overdamped Langevin equation
\begin{equation} \frac{dx(t)}{dt}=\sqrt{2D(t)}\times\zeta(t)\label{eq-over} \end{equation} driven by zero-mean and unit-variance Gaussian noise $\zeta(t)$. At step $i+1$ the particle displacement is given by \begin{equation} x_{i+1}-x_i=\sqrt{2[D(t_i)+D_0]}\times(y_{i+1}-y_{i}), 
\label{eq-simul-scheme}\end{equation} where $(y_{i+1}-y_i)$ are the increments of the Wiener process. Unit time intervals
separate consecutive steps. To avoid possible particle stalling we regularise $D$ by adding a small constant $D_0=10^{-3}$ \cite{cher14a,cher14b}. This does not affect the intermediate- and long-time diffusive behaviour. The particle's initial position is $x_0=x(t=0)=0.1.$ In the second part of the paper, we simulate the underdamped Langevin equation for a particle of
mass $m$,
\begin{equation}
m\frac{d^2x(t)}{dt^2}+\eta\frac{dx(t)}{dt}=\sqrt{2D(t)}\times\zeta(t),
\label{eq-under}
\end{equation}
with the unit damping coefficient set below to $\eta=1$. At a step $i+1$ the particle displacement is found from the iteration scheme
\begin{eqnarray} 
\nonumber
&m(x_{i+1}-2x_i+x_{i-1})+\eta(x_{i+1}-x_i)\\
&=\sqrt{2[D(t_i)]}\times(y_{i+1}-y_{i}),
\label{eq-simul-scheme-massive}
\end{eqnarray}
where the instantaneous diffusivity is taken from Eq.~(\ref{eq-pD-Rayleigh}) with
the mean (\ref{eq-sigma-power-law}). 

\begin{figure*}
\includegraphics[width=18cm]{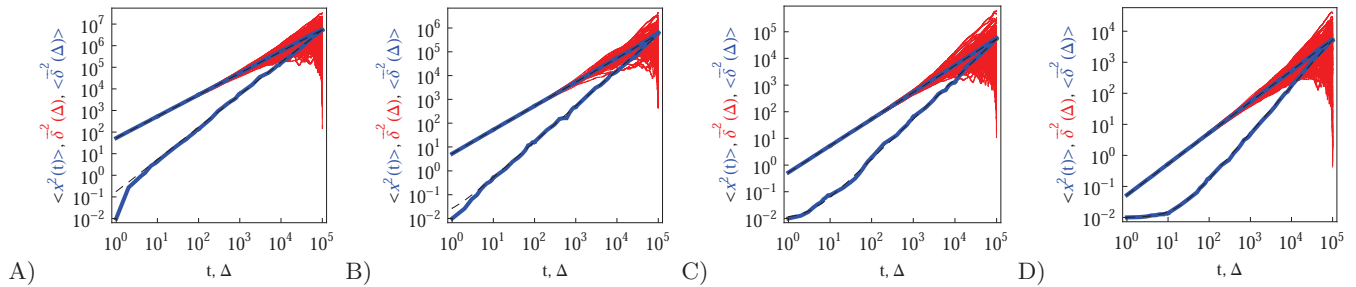}
\caption{MSD $\left<x^2(t)\right>$ and time averaged MSD $\left<\overline{\delta
^2(\Delta)}\right>$ (thick blue curves) as well as individual time traces $\overline{
\delta^2(\Delta)}$ (red curves) for overdamped RDPs. The asymptotes
(\ref{eq-msd-magnitude}) and (\ref{eq-tamsd-magnitude}) for the MSD and the time
averaged MSD are shown by the dashed curves. The asymptotes often superimpose with 
the results of simulations. Parameters: scaling exponent is $\omega=1/2$, 
trace length is $T=10^5$, and number of traces used for the averaging is
$N=150$. The starting position is $x_0=0.1$ and the parameter $2\gamma\mathscr{D}/
\sqrt{2\pi}$ takes the values $10^{-1}$, $10^{-2}$, $10^{-3}$, and $10^{-4}$ for
the panels from left to right. The variable of the $x$ axis is the time $t$ or the lag
time $\Delta$.}
\label{fig-msd-tamsd}\end{figure*}

\section{Results: Overdamped Motion}
\label{sec-results-over}

\subsection{Free Diffusion}

We start with the diffusion of massless particles with a fluctuating diffusivity and time invariant mean. As naively
expected, we find that due to friction the MSD in the long time limit is Brownian,
\begin{equation}
\left<(x^2(t)\right>\sim x_0^2+2\left<D(t)\right>t.
\label{eq-msd}
\end{equation}
The MSD and the time averaged MSD are nearly identical after a fast relaxation
of the starting position $x_0$ of the particle and the ergodicity is
approximately fulfilled at all times (results not shown here).

Now we address the more interesting case of RDPs with instantaneous diffusivity
chosen from the distribution (\ref{eq-pD-Rayleigh}) with a time dependent width
(\ref{eq-sigma-power-law}). Namely, the most likely diffusivity $D_\sigma(t)$ at
simulation step $i$ is chosen as
\begin{equation}
(D_\sigma)_i=\frac{2\gamma\mathscr{D}}{\sqrt{2\pi}}i^{\omega},
\label{eq-sigma-time-dependence}
\end{equation}
where the prefactor is chosen for convenience and the
coefficient $\mathscr{D}$ tunes the magnitude of the diffusivity. From
Eq.~(\ref{eq-over}) we straightforwardly compute
\begin{equation}
\left<x^2(t)\right>=x_0^2+2\mathscr{D}t^{\gamma},
\label{eq-msd-magnitude}
\end{equation}
with
\begin{equation}
\gamma=1+\omega.
\end{equation}

Fig.~\ref{fig-msd-tamsd} depicts the case of $\omega=1/2$ or the MSD diffusion exponent $\gamma=3/2$. As can be
seen from the simulations the time averaged MSD grows linearly with the
lag time $\Delta$, as in the Brownian case. This process also reveals
a quite moderate amplitude spread of individual traces $\overline{\delta^2(\Delta)}$, 
see the thin red curves in Fig.~\ref{fig-msd-tamsd}.
Obviously when the lag time approaches the observation time, $\Delta\sim T$, the
amplitude scatter increases due to the deteriorating statistic of $\overline{\delta
^2(\Delta)}$ \cite{metz14}. Varying $\mathscr{D}$ in Fig.~\ref{fig-msd-tamsd} we demonstrate that, as expected, larger
initial diffusivities give rise to a faster approach of the MSD to the
theoretical asymptote. In contrast, for rather small diffusivities
(smaller $\mathscr{D}$ values) the system needs more time to approach the long
time asymptote, Fig.~\ref{fig-msd-tamsd}. A diminished magnitude of the MSD at smaller $\mathscr{D}$ values inevitably leads
to a decrease in the magnitude of the time averaged MSD, see below. At intermediate
to long times the superdiffusive MSD regime with $\gamma=3/2$ emerges. Finally, towards
the very end of the trace the MSD and the time averaged MSD coincide, as they should \cite{metz14}.

Analytically, we obtain for the time averaged MSD from Eq. (\ref{eq-over})
for $x(t)$ and after averaging over the noise $\zeta(t)$ and diffusion coefficient realisations $p(D)$ the result
\begin{equation}
\left<\overline{\delta^2(\Delta)}\right>=\frac{2\mathscr{D}\left[T^{\gamma+1}
-(T-\Delta)^{\gamma+1}-\Delta^{\gamma+1}\right]}{(\gamma+1)(T-\Delta)}.
\label{eq-tamsd-magnitude}
\end{equation}
This expression nicely agrees with the results of computer
simulations for all $\omega$ and $\mathscr{D}$ values investigated, see Figs.~\ref{fig-msd-tamsd} and
\ref{fig-msd-tamsd-omega}. It is not surprising that both MSD and time averaged
MSD are proportional to $\mathscr{D}$ determining the basal value of the particle
diffusivity, Eq. (\ref{eq-sigma-time-dependence}). In the limit of short
lag times, $\Delta\ll T$, from  Eq.~(\ref{eq-tamsd-magnitude}) we recover the
scaling behaviour
\begin{equation}
\left<\overline{\delta^2(\Delta)}\right>\simeq\frac{2\mathscr{D}\Delta}{
T^{1-\gamma}}.
\label{eq-t-dependence-ala-hdps}
\end{equation}
Thus, for superdiffusive RDPs with $\gamma>1$ the magnitude of the time averaged
MSD is a growing function of the trace length $T$, while for subdiffusive RDPs 
$\left<\overline{\delta^2}\right>$ magnitude decreases with $T$,
in agreement with Fig.~\ref{fig-tamsd-scaling-t}. In the limit $\Delta\to T$ the
ensemble and time averaged MSDs coincide \cite{metz14}, as it is easy to check
from Eqs.~(\ref{eq-msd-magnitude}) and (\ref{eq-tamsd-magnitude}) and corroborated
in Fig.~\ref{fig-msd-tamsd}. The non-equivalence of the time averaged MSD
(\ref{eq-tamsd-magnitude}) and the time averaged MSD (\ref{eq-msd-magnitude})
demonstrates that the system is weakly non-ergodic. The scaling behaviour 
(\ref{eq-t-dependence-ala-hdps}) regarding the magnitude of
the time averaged MSD in terms of the power law of the trace length $T$ and the
linearity in the lag time is analogous to that obtained for subdiffusive
continuous time random walks \cite{metz10,metz11b,metz08,soko08} and their
correlated version \cite{corr1} as well as for HDPs \cite{cher13,cher14a,cher15b}
and SBM \cite{jeon14,jeon15}; see also Ref. \cite{metz14} for overview.

For RDPs the ergodicity breaking parameter EB computed from simulations tends to
follow the asymptote (\ref{eq-eb-bm}) for Brownian motion at intermediate
and long times, Fig.~\ref{fig-eb}. We observe that the initial 
relaxation of EB to this asymptote is relatively fast. 
As we show in Fig.~\ref{fig-eb}, after this relaxation time
the parameter $\mathcal{EB}$ (\ref{eq-eb2}) becomes a power-law
function of the lag time $\Delta$,
\begin{equation}
\mathcal{EB}(\Delta)\simeq\Delta^{1-\gamma}.
\label{eq-eb2-scaling}
\end{equation}

\begin{figure}
\includegraphics[width=6.8cm]{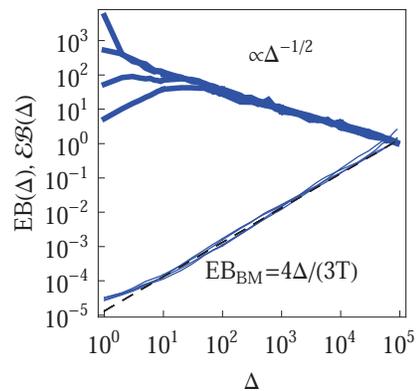}
\caption{Ergodicity breaking parameter EB and the parameter $\mathcal{EB}$ versus
lag time $\Delta$ for overdamped RDPs. The curves are computed for the parameters
of Fig.~\ref{fig-msd-tamsd} with $\omega=1/2$. The curves for $\mathcal{EB}$ from
top to bottom correspond to $2\gamma\mathscr{D}/\sqrt{2\pi}=10^{-1}$, $10^{-2}$,
$10^{-3}$, and $10^{-4}$, respectively. The asymptote (\ref{eq-eb-bm}) of 
Brownian motion and the relation (\ref{eq-eb2-scaling}) are the dashed
lines.}
\label{fig-eb}
\end{figure}

\begin{figure}
\includegraphics[width=6.8cm]{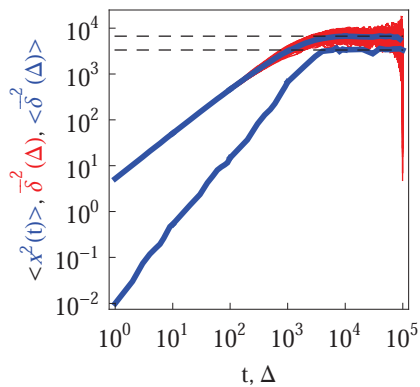}
\caption{Same as in Fig.~\ref{fig-msd-tamsd} but for confined overdamped RDPs.
The box size is $L=10^2$,  $\omega=1/2$, and $2\gamma\mathscr{D}/\sqrt{2\pi}=0.01$. The asymptote (\ref{eq-msd-confined})
and twice this value are the dashed lines.}
\label{fig-msd-tamsd-confined}
\end{figure}

\begin{figure}
\includegraphics[width=6.8cm]{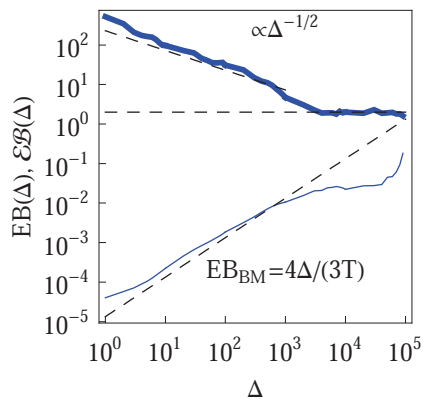}
\caption{Parameters EB and $\mathcal{EB}$ for confined overdamped RDPs for the parameters of Fig.~\ref{fig-msd-tamsd-confined}.
The Brownian asymptote (\ref{eq-eb-bm}) and the plateau $\mathcal{EB}=2$ are shown as the dashed lines.}
\label{fig-eb-confined}
\end{figure}

Diffusivity distributions $p(D)$ whose mean diffusivity grows with
time may be viewed to correspond to an effectively increasing temperature in the
system. The opposite case of a temporally shrinking width 
may stem from a cooling of the system in the course of time. In this respect the
current process is reminiscent of SBM \cite{jeon14,safd15,jeon15}. An important
example for the latter are granular gases with a relative velocity dependent
restitution coefficient \cite{bodr15gg}.

The particle spreading for very subdiffusive RDPs can be compared to the PDF of
SBM, identical to that of fractional Brownian motion for $x_0=0$ \cite{jeon14,muni02}.
Namely, after the substitution of the corresponding MSD
(\ref{eq-msd-magnitude}) this produces \begin{equation}
P(x,t)\sim\frac{1}{\sqrt{4\pi\mathscr{D}t^{\gamma}}}\exp\left(-\frac{x^2}{4\mathscr{
D}t^{\gamma}}\right). \label{eq-pdf-ala-sbm} \end{equation}
Fig.~\ref{fig-pdf-spiky} compares the conjecture (\ref{eq-pdf-ala-sbm}) with the
result from simulations of RDPs. We see, however, that as
the MSD scaling exponent increases a distinct cusp of the PDF starts to develop
at the origin. The PDF of superdiffusive RDPs becomes pronouncedly non-Gaussian.
This spike cannot be described by a convolution of the
diffusivity distribution (\ref{eq-pD-Rayleigh}) with the kernels of Brownian motion
and SBM motion. A more detailed investigation is required to understand this spike at short times and possibly exponential forms of the PDF tails at long times. 
This generally non-Gaussian and $\alpha$-dependent PDF shape is one important distinction of RDPs with time dependent \textit{and in addition} fluctuating diffusivities, as compared to the SBM process with the same diffusion coefficient value at each step, $D(t)=\left< D(t) \right>$.

\subsection{Confined Diffusion}

We now turn to confined RDPs on an interval $-L<x<L$. Such confined motion is
important especially for the understanding of diffusion processes in biological
cells. In cells, due to their external confinement by the plasma membrane and 
internal compartmentalisation a diffusing tracer frequently collides with boundaries. 
As expected, after an initial free
diffusion the MSD converges to the stationary plateau \cite{metz14}
\begin{equation}
\left<x_{\text{st}}^2\right>=\frac{1}{2}\left<\overline{\delta^2_{\text{st}}}
\right>=\frac{1}{3}L^2,
\label{eq-msd-confined}
\end{equation}
as demonstrated in Fig.~\ref{fig-msd-tamsd-confined}. The time averaged MSD,
by virtue of its definition (\ref{eq-tamsd}), approaches twice the value of the
MSD in the long time limit. The existence of a plateau is similar to that of
standard Brownian motion, and interval-confined SBM and HDPs \cite{cher14b,
cher15b}. Note that SBM confined by an external potential has a time dependent
thermal value of the MSD \cite{jeon14,jeon15}. The behaviour of confined continuous time
random walks is strikingly different, there confinement leads to a crossover to
a second power-law regime in the time averaged MSD \cite{metz10,metz11b}. The PDF of confined RDPs approaches a uniform distribution of particles on the interval.

At more severe confinement the ergodicity breaking EB parameter at large lag times
$\Delta$ values starts to deviate form the Brownian asymptote (\ref{eq-eb-bm}), see
Fig.~\ref{fig-eb-confined}. In addition, we find that for a fixed width of the
confining interval and varying trace length $T$ the EB parameter follows the
scaling relation \begin{equation} \textrm{EB}(T)\simeq1/T,\label{eq-eb-versus-t}
\end{equation} as illustrated in Figs.~\ref{fig-eb-confined-t}A,B. This is a standard decrease
of the EB parameter for longer trajectories, a property ubiquitous among a number of both ergodic and non-ergodic stochastic processes
\cite{metz14}. The decrease of $\mathrm{EB}$ with $T$ indicates a progressively 
more ergodic diffusion for longer particle traces. 

\begin{figure*}
\includegraphics[width=18cm]{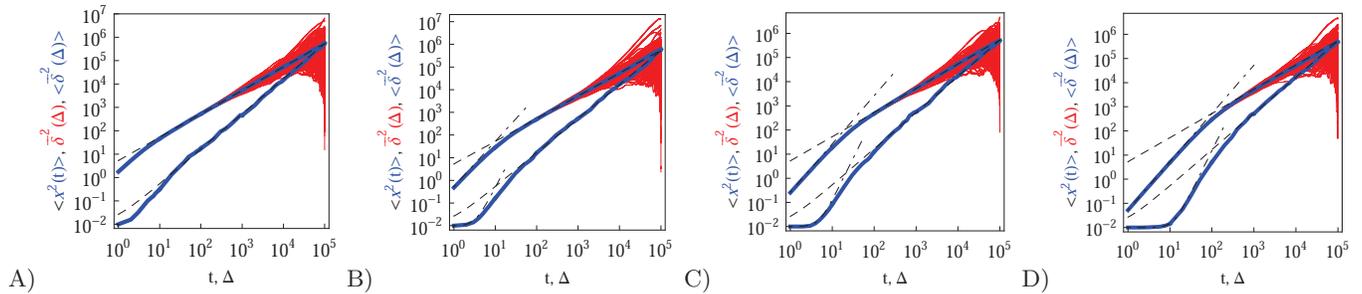}
\caption{Ensemble averaged MSD, individual time averaged MSDs, and mean time
averaged MSD, with the same notation as in Fig.~\ref{fig-msd-tamsd}, evaluated
for underdamped RDPs with particle mass $m=1$, $5$, $10$, and $50$, for panels
from left to right. $N\sim10^2$ traces are shown. The long time asymptotes
(\ref{eq-msd-magnitude}) and (\ref{eq-tamsd-magnitude}) are the dashed lines. 
The dot-dashed lines are the asymptotes for the short
time regimes, Eqs. (\ref{eq-initial-ballistic-msd}) and
(\ref{eq-initial-ballistic-tamsd-large-beta}). Parameters: $2\gamma\mathscr{
D}/\sqrt{2\pi}=0.01$, $\omega=1/2$, and $T=10^5 $.}
\label{fig-msd-tamsd-underdamped}
\end{figure*}

\begin{figure}
\includegraphics[width=6.8cm]{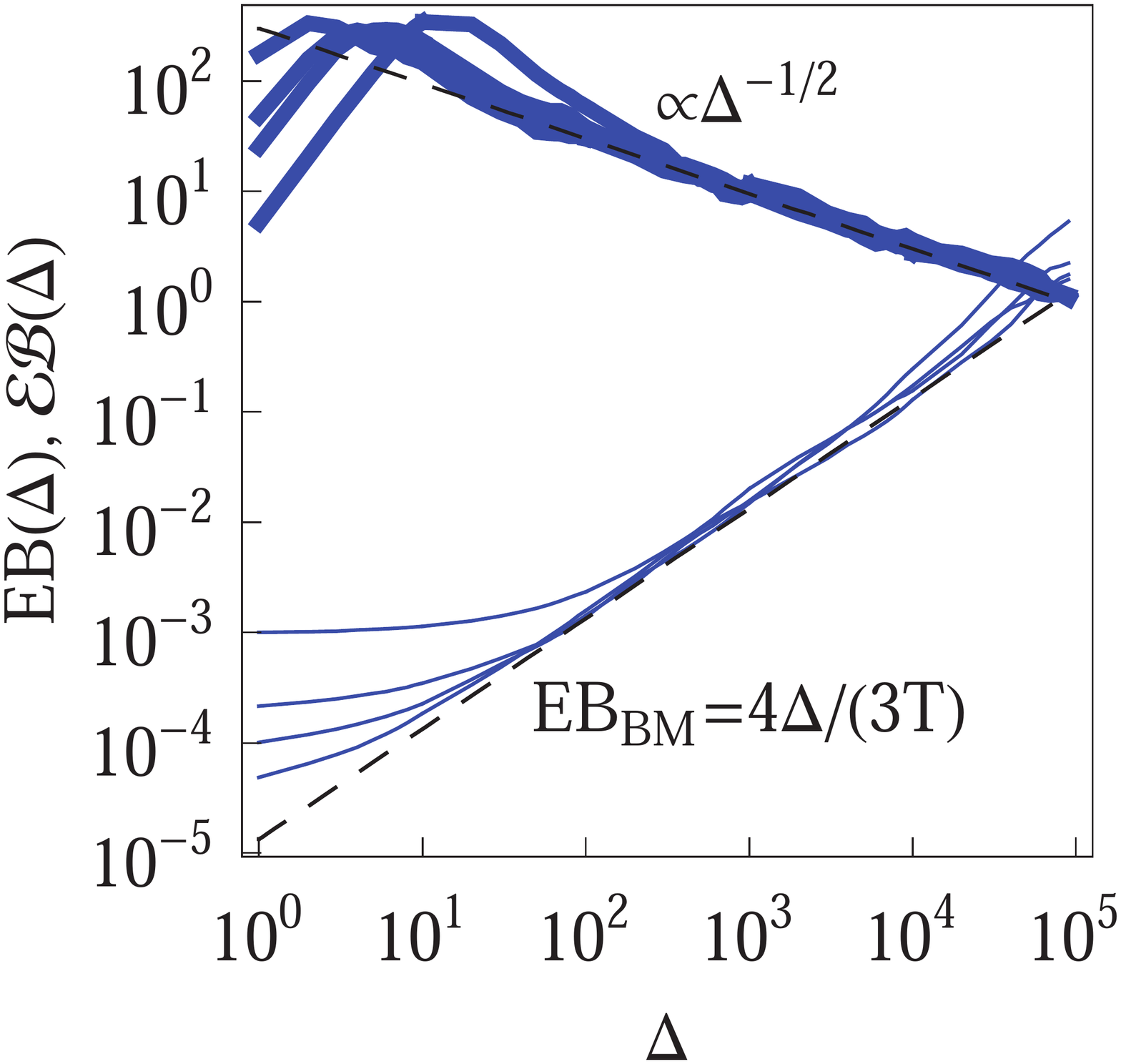}
\caption{Parameters EB and $\mathcal{EB}$ versus lag time $\Delta$ for
underdamped RDPs. The Brownian asymptote (\ref{eq-eb-bm}) and Eq. (\ref{eq-eb2-scaling}) are the dashed lines. 
The curves correspond to varying particle mass, $m=1$, $5$, $10$, $50$; the other parameters are the
same as in Fig.~\ref{fig-msd-tamsd-underdamped}.}
\label{fig-eb-under}
\end{figure}

\section{Results: Underdamped Motion}
\label{sec-results-under}

\subsection{Free Diffusion}

In this section, we study the diffusion of massive particles in the same time
dependent random diffusivity scenario (\ref{eq-sigma-time-dependence}) based on the underdamped Langevin equation. 
In particular, we explore to what extent inertia effects modify the long time behaviour of the MSD and the time averaged
MSD, as compared to the overdamped RDPs considered above. 
The general solution for the particle MSD follows from the standard procedure for
the Brownian motion of massive particles \cite{uhle30}. Namely, we obtain
\begin{eqnarray}
\nonumber
x(t)-x(0)&=&\frac{v_0}{\theta}\left(1-e^{-\theta t}\right)\\
\nonumber
&&-\frac{A}{\theta}e^{-\theta t}\int_0^t(t')^{\omega/2}\zeta(t')e^{\theta t'}
dt'\\
&&+\frac{A}{\theta}\int_0^t(t')^{\omega/2}\zeta(t')dt',
\label{eq-xt}
\end{eqnarray}
where we defined $\theta=\eta/m$ and $A^2=2\gamma\mathscr{D}/m^2$.
Moreover, $v_0$ is the initial particle velocity. Then the MSD of the particles
after averaging over the noise $\zeta$ can formally be written as
\begin{eqnarray}
\nonumber
\left<x^2(t)\right>&=&x_0^2+\frac{v_0^2}{\theta^2}\left(1-e^{-\theta t}\right)^2\\
\nonumber
&&-2\gamma\mathscr{D}e^{-2\theta t}\Big[t^{\gamma}\mathrm{Ei}(1-\gamma,-2\theta
t)\Big]_0^t+2\mathscr{D}t^{\gamma}\\
&&+4\gamma\mathscr{D}e^{-\theta t}\Big[t^{\gamma}\mathrm{Ei}(1-\gamma,-\theta
t)\Big]_0^t.
\label{eq-msd-underdamped}
\end{eqnarray}
Here $\mathrm{Ei}(n,z)=\int_1^{\infty}e^{-zt}t^{-n}dt$ denotes the generalised
exponential integral.


For zero initial velocity of the particles $v_0=0$, as in the computer simulations performed here, the inertial term in the Langevin equation gives rise to the initial MSD scaling of the form $\left<x^2(t)\right>\sim t^{\gamma+2}$. It is due to progressively accelerating (heating) particles. Explicitly, for the MSD at short times we get \begin{equation} \left< x^2(t)\right>\sim x_0^2+\frac{4\mathscr{D}\eta^2t^{\gamma+2}}{m^2(\gamma+
1)(\gamma+2)}. \label{eq-initial-ballistic-msd}\end{equation}
This faster than ballistic MSD regime often called hyperdiffusion is known to
emerge, for instance, for a power-law like transient heating of particles with
temperature variation of the form $\mathcal{T}(t)\simeq t^\omega$ \cite{goyc10}. 
This superballistic behaviour emerges for RDPs with fluctuating diffusivities and time dependent mean, in analogy with a faster than linear short time ballistic regime in Brownian motion \cite{uhle30}.

Note that this short time MSD regime---with the scaling exponent by one larger than the long time exponent---is absent in the model of underdamped scaled Brownian motion elucidated by us recently in Ref.  \cite{bodr15gg}. The reason is that the damping coefficient is set to be temperature independent in the current model, whereas in the model of underdamped SBM the parameter $\eta(t)$ is coupled to law of diffusivity variations via the generalised time-local Einstein relation \begin{equation} D(t)=k_B T(t)/(m\eta(t)) \label{einstein} \end{equation} \cite{bodr15gg}. So, the fluctuation-dissipation theorem is valid, contrary to the current approach. For the underdamped SBM process, the relation $\eta(t) \sim T^{1/2}(t)$ is consistent with the physical picture of elastically colliding and relaxing particles in a bath with a deterministically varying temperature \cite{bodr16}. The reader is also referred to \cite{lind07} for studying different relationships between the friction coefficient and velocity for passive and active \cite{roma12} particles, including the nonlinear forms.

Note that the diffusive and ergodic properties of underdamped SBM were recently considered too \cite{bodr16}. It was demonstrated that the inertial effects relax rather quickly in the course of particle diffusion for $\alpha>1$ situations. This follows from comparing the magnitudes of the acceleration and frictional terms in the Langevin equation. On the other hand, for small positive values of $\alpha$ a finite particle mass yields an extensive intermediate regime, both for the MSD and the time averaged MSD growth behaviour with time. Interestingly, in the case of ultraslow logarithmic SBM motion realised at the boundary value $\alpha=0$ the overdamped limit of particle diffusion \cite{bodr15} is not reached at long times, independent on the total measurement time \cite{bodr16}.

The time averaged MSD of underdamped RDPs follows from Eqs.~(\ref{eq-xt}) and (\ref{eq-tamsd}), 
\begin{eqnarray}
\nonumber
\left<\overline{\delta^2(\Delta)}\right>&=&\frac{A^2\eta^2}{\theta^2(T-\Delta)}
\int_0^{T-\Delta}dt\left\{\frac{(t+\Delta)^{\gamma}}{\gamma}-\frac{(t)^{\gamma}}{
\gamma}\right.\\
\nonumber 
&&\hspace*{-1.2cm}-e^{-2\theta (t+\Delta)}\left[t^{\gamma}\mathrm{Ei}(1-\gamma,-2
\theta t)\right]_0^{t+\Delta}\\
\nonumber
&&\hspace*{-1.2cm}-\left(e^{-2\theta t}-2e^{-\theta t}e^{-\theta (t+\Delta)}\right)
\left[t^{\gamma}\mathrm{Ei}(1-\gamma,-2 \theta t)\right]_0^t\\
\nonumber
&&\hspace*{-1.2cm}+2e^{-\theta (t+\Delta)}\left[t^{\gamma}\mathrm{Ei}(1-\gamma,-
\theta t)\right]_0^{t+\Delta}\\
&&\hspace*{-1.2cm}-\left. 2e^{-\theta (t+\Delta)}\left[t^{\gamma}\mathrm{Ei}(1-
\gamma,-\theta t)\right]_0^t\right\}.
\label{eq-tamsd-underdamped}
\end{eqnarray}
This integral expression can be evaluated numerically. In the limit of short lag
times $\Delta \ll T$ we can evaluate the integral and find
\begin{eqnarray}
\nonumber
\left<\overline{\delta^2(\Delta)}\right>&\sim&\gamma\mathscr{D}\frac{e^{-2\theta T}
(-1)^{1-\gamma}}{2(2\theta)^{\gamma-1}}\\
&&\hspace*{-0.8cm}\times\left[\Gamma(\gamma+1)-\Gamma(\gamma+1,-2\theta T)\right]
\frac{\Delta^2}{T}.
\label{eq-initial-ballistic-tamsd}
\end{eqnarray}
Here $\Gamma(a,x)=\int_x^{\infty}t^{a-1}e^{-t}dt$ is the generalised incomplete
Gamma function and $\Gamma(a,0)=\Gamma(a)$ is the Gamma function. The short time
asymptotes of both the MSD and the time averaged MSD are plotted as the dot-dashed
curves in Fig.~\ref{fig-msd-tamsd-underdamped} showing nice agreement with the
results of computer simulations of Eq.~(\ref{eq-simul-scheme}).

Expanding the Gamma functions in the corresponding limit we find from expression
(\ref{eq-initial-ballistic-tamsd}) for light particles or high friction in the
system---that is for $\theta T\gg1$---that
\begin{eqnarray} \left<\overline{\delta^2(\Delta)}\right> \sim\frac{\mathscr{D}\eta T^{\gamma-1} \Delta^2}{m} \sim\frac{D(T)\eta\Delta^2}{m\gamma}.
\label{eq-initial-ballistic-tamsd-large-beta} \end{eqnarray} 
This has the form of the short time MSD behaviour of standard Brownian motion \cite{uhle30}. All particle masses in this study guarantee the validity of
Eq.~(\ref{eq-initial-ballistic-tamsd-large-beta}) as the short time expansion for
the time averaged MSD in this underdamped limit. Comparing the $\overline{\delta^2(T)}$-variation with Eq.~(\ref{eq-initial-ballistic-tamsd-large-beta}) in 
Fig.~\ref{fig-tamsd-scaling-t}B further supports this validity regarding the dependence on the length of particle trajectory $T$. In the opposite limit of $\theta T\ll1$---very massive particles or low friction for the particle motion---we find \begin{eqnarray}
\left<\overline{\delta^2(\Delta)}\right>\sim\frac{2\mathscr{D}\eta^2 T^{\gamma}
\Delta^2}{m^2(\gamma+1)}\sim\frac{\Delta^2}{m^2}.
\label{eq-initial-ballistic-tamsd-small-beta}
\end{eqnarray}

\begin{figure*}
\includegraphics[width=18cm]{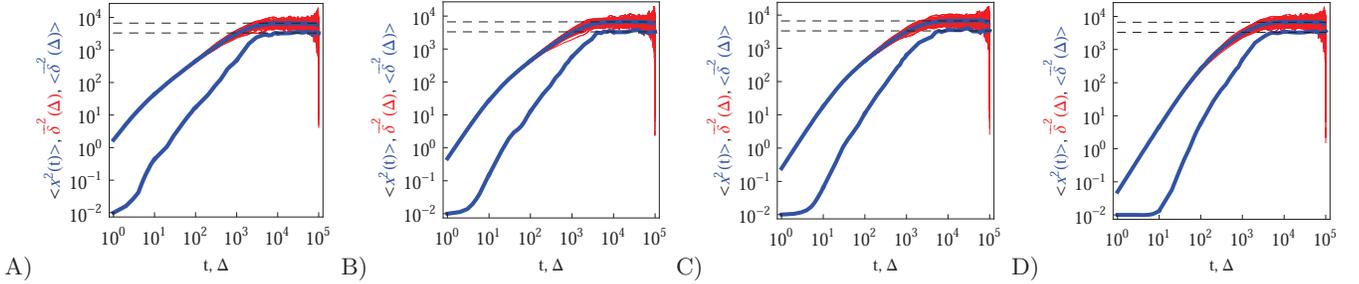}
\caption{Analogue of Fig.~\ref{fig-msd-tamsd-underdamped} for confined underdamped
RDPs with particle mass $m=1$, $5$, $10$, and $50$, from left to right. We show
$N=150$ traces for each set of parameters and the interval width is $L=10^2$.
The asymptotes for the confined motion (\ref{eq-msd-confined}) are the dashed lines.}
\label{fig-msd-tamsd-underdamped-confined}
\end{figure*}

Let us now describe the results of our computer simulations and compare them
with these analytical predictions. We observe that in the particle displacement
the initial condition of zero particle velocity ($v_0=0$) relaxes within several
initial diffusion steps. Naturally, it takes for the system longer to accelerate
heavier particles, as shown in Fig. \ref{fig-msd-tamsd-underdamped} in agreement
with Eq. (\ref{eq-initial-ballistic-msd}). The initial slow acceleration of heavy
particles yields slowly growing MSD that gets in turn reflected in small amplitudes
of the time averaged MSD at short lag times $\Delta$. The MSD follows Eq. (\ref{eq-initial-ballistic-msd}) for short times and then crosses over to the long time scaling (\ref{eq-msd-magnitude}). Note that the same initial quadratic regime was observed in Ref. \cite{bodr16} for the short time behaviour of the time averaged MSD of the standard underdamped SBM process. For the time averaged MSD the heavier particles feature a longer ballistic regime, see Fig. \ref{fig-msd-tamsd-underdamped}. The magnitude of $\left< \overline{\delta^2} \right>$ decreases with the particle mass, in agreement with Eq. (\ref{eq-initial-ballistic-tamsd-large-beta}). At longer times the MSD and time averaged MSD approach the results expected for the overdamped RDP motion, shown as the long time asymptotes in Fig. \ref{fig-msd-tamsd-underdamped}. As expected, apart from the initial ballistic regime of the time averaged MSD described by Eq. (\ref{eq-initial-ballistic-tamsd-large-beta}), the analytical solution for the overdamped limit (\ref{eq-tamsd-magnitude}) describes the long time behavior of the results of our computer simulations.

Similar to the overdamped situation, at short lag times the magnitudes of 
 $\left<\overline{\delta^2}\right>$ follow the relation (\ref{eq-t-dependence-ala-hdps}),
see Fig.~\ref{fig-tamsd-scaling-t}B. We also observe that for underdamped RDPs
the spread of individual $\overline{\delta^2}$ traces is similar to that of
standard Brownian motion. This small spread is consistent with the observation
that the EB parameter for this underdamped RDP motion does
not deviate strongly from the Brownian asymptote (\ref{eq-eb-bm}) at intermediate
and long times, see Fig.~\ref{fig-eb-under}. In the region of short $\Delta$ the
deviations are quite substantial, particularly for massive particles exhibiting a
ballistic initial growth of $\left<\overline{\delta^2}\right>$ and 
a nonlinear growth of the MSD, see Fig.~\ref{fig-msd-tamsd-underdamped}. The
auxiliary parameter $\mathcal{EB}$---again after the initial particle acceleration---follows the asymptote (\ref{eq-eb2-scaling}), see the thick curves in Fig.~\ref{fig-eb-under}. The deviations of EB and $\mathcal{EB}$ at short lag times from the Brownian asymptote is more evident for massive particles.

\subsection{Confined Diffusion}
 
We complete our analysis of RDPs with the study of the underdamped motion in
a confining box. We observe that at short time the MSD develops similar to the
unconfined scenario. Once the boundary of the confined region is reached,
the plateaus start to develop at the same levels as for the overdamped RDP
case both for the MSD and time averaged MSD, see Eq.~(\ref{eq-msd-confined})
and Fig.~\ref{fig-msd-tamsd-underdamped-confined}. Towards the very end of the
trajectory at $\Delta=T$, the MSD and the time averaged MSD coincide, as they
should \cite{metz14}. The longer the entire trajectory, however, the narrower the range of lag times
where this convergence takes place, and thus the more precise should be the
$\Delta$-sampling in this region that is often computationally costly. This
effect was studied in detail for the pure SBM motion confined in harmonic
potentials \cite{jeon14} and for HDPs confined between the hard walls \cite{cher14b,cher15b}.

The PDFs of confined underdamped RDPs at varying box width $L$ is 
presented in Fig.~\ref{fig-pdf-confined-l}. We observe that for wide 
 intervals the particles are nearly uniformly distributed on the interval (see
the dashed lines in Fig.~\ref{fig-pdf-confined-l}), with only insignificant
increase in the particle occupancies near the box boundaries due to reflections.
Note that the particle starting position at $x=x_0$ is still slightly visible in
the PDF for a weak confinement. As the confinement becomes more severe,
the particle accumulation bear the interval boundaries occupies a larger
fraction of space available for diffusion.

The EB parameter for massive particles deviates progressively from the Brownian
law (\ref{eq-eb-bm}) at both short and long lag times $\Delta$ (not shown).
This is due to slow particle acceleration at short times (MSD plateau) and particle confinement at long
times, respectively. For free and confined underdamped RDPs---similarly to the
overdamped situation---the EB parameter follows the asymptote (\ref{eq-eb-versus-t})
with the trajectory length $T$, see Figs.~\ref{fig-eb-confined-t}C,D. As the
confinement becomes less severe, the EB parameter approaches the
value for the free underdamped RDP motion. Fig.~\ref{fig-eb-confined-l}
illustrates this EB evolution with the width of the confining interval.

\section{Conclusions}
\label{sec-disc}

We examined the ensemble and time averaged characteristics of random diffusion
processes. The randomness of the diffusion coefficient $D$ was implemented in
the model via a non-stationary distribution $p(D)$ of diffusivities.
RDPs are not thermalised, that is, the motion of the particles is inherently out
of equilibrium. The distribution of the diffusion coefficient reflects individual
variability of particles diffusivities and heterogeneities of the environment. 
For typical out-of-equilibrium systems such as biological cells this does not pose
any restrictions to our model. RDPs represent a quite flexible model to study
asymptotically Brownian and anomalous diffusive systems with a locally fluctuating
diffusion coefficient to model physical situations in many complex systems.

We computed both by computer simulations and analytically the MSD, the time
averaged MSD, and the ergodicity breaking parameter of RDPs. 
The unconfined and interval-confined motion were examined. We found that in terms of these
standard characteristics subdiffusive RDPs appear similar to subdiffusive SBM with
a deterministic diffusivity variation in time. For superdiffusive RDPs the
fluctuations grow with time. Concurrently, the average diffusion coefficient $\left<D(t)\right>$ grows with time together with the spread of its values. These
features reflect an increasing temperature and more pronounced fluctuations
of the medium in the course of particle diffusion. The properties of ageing overdamped and underdamped RDPs will be considered elsewhere.

Living cells feature heterogeneous and densely crowded environments established by a melange of various macromolecules and (importantly) a rather viscoelastic solution between them. This often leads to a broadening in the distribution of diffusion coefficients and subdiffusive exponents, as observed for obstructed diffusion of various tracers \cite{saxt97,saxt01,webb96,weis04}. In particular, some extensions of the standard diffusion models to account for these effects---similar to our $p(D)$ distribution for the SBM like diffusion model presented above---appear necessary e.g. for a quantitative fit of fluorescence recovery after photobleaching curves \cite{saxt01}. The models of SBM type---with the diffusivity formally decaying in time according to the power law (\ref{eq-sigma-power-law})---are often implemented to describe the subdiffusive MSD behavior (\ref{eq-ad}) of the tracer particles in cells. This anomalous MSD scaling was observed e.g. via fitting the shape of the autocorrelation curves of fluorescence correlation spectroscopy measurements \cite{webb99,weis04,weis03}. In single particle tracking measurements in biological cells some time dependent scatter of the diffusion coefficients was also detected  \cite{saxt97, kusu93}. It is necessary in theoretical models i.a. to distinguish between the normal, restricted, and fully trapped populations of the tracers. In fact, a Gamma distribution similar to the Rayleigh distribution (\ref{eq-pD-Rayleigh}) used above was proposed in Ref. \cite{saxt97} to characterise the scatter of the MSD's distribution of diffusing particles. On the level of diffusing simple organisms, Gamma like diffusivity distributions were documented for the motion of nematode worms \cite{hapc09}. The latter also exhibit non-Gaussian PDFs of the particle displacements with a "spike" at the origin \cite{hapc09,hapc07}, similar to some of our findings. Also, the recent study \cite{para16} of anomalous and non-ergodic dynamics of particles within a predator-prey model with a broad distribution of diffusion coefficients of interacting partners needs to be mentioned here.

The current study with its preset functional form of the diffusivity distribution and a deterministic law (\ref{eq-sigma-power-law}) represents a first step into the terrain of stochastic processes with fluctuating and time varying diffusivities. A more general consideration would correspond to a system of coupled stochastic differential equations for the particle position and its diffusivity. The first equation is the standard Langevin equation, while the second equation involves an additional, generally decoupled noise source governing $D(t)$ variation. The correlation function and other noise properties---not necessarily Gaussian---determine then both the ensemble and the time averaged MSDs of diffusing particles \cite{cher17}.

\numberwithin{figure}{section}
\renewcommand{\thefigure}{A\arabic{figure}}

\appendix

\section{}

In this Appendix we present several additional figures supporting the claims in
the main text of the manuscript.

\begin{figure}
\includegraphics[width=8.8cm]{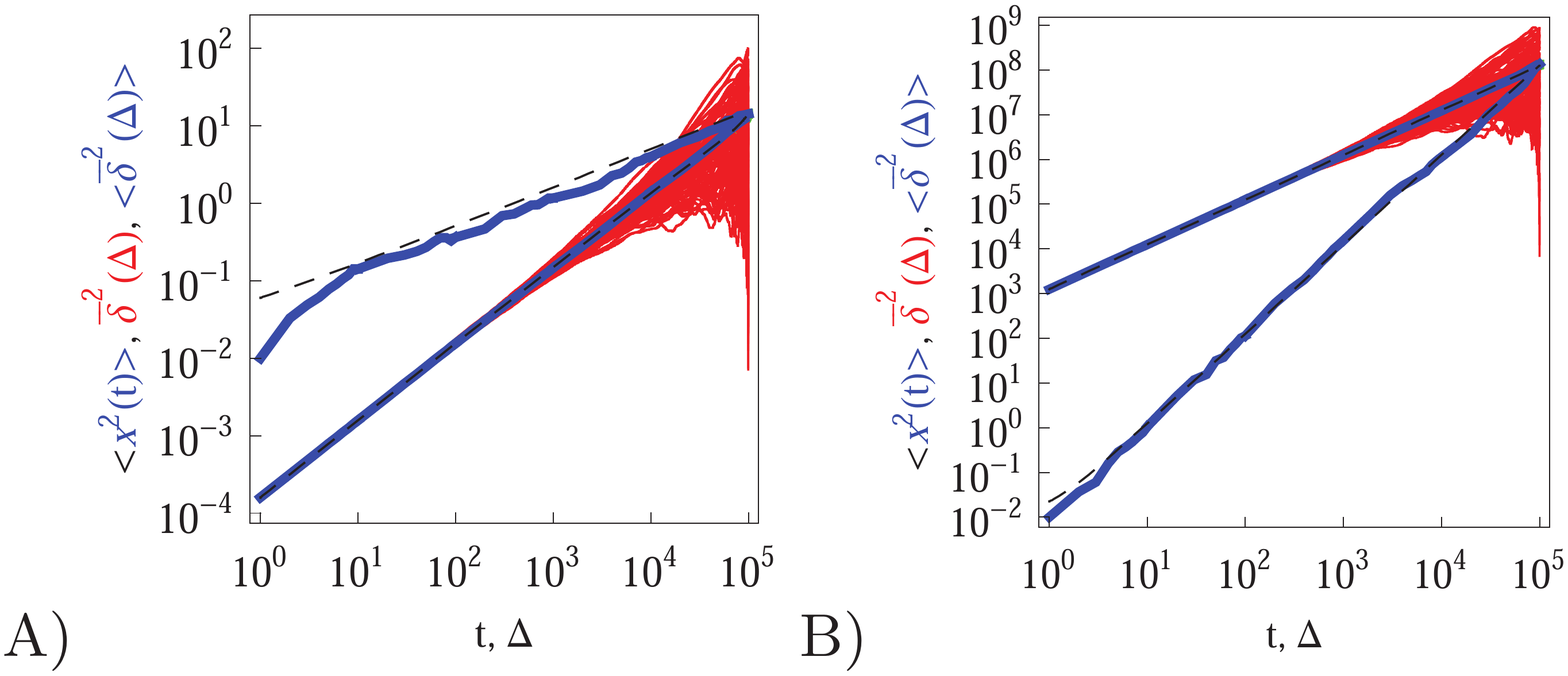}
\caption{Same as in Fig. \ref{fig-msd-tamsd} but at $\omega=
-1/2$ (A) and $\omega=1$ (B) computed for $2\gamma\mathscr{D}/\sqrt{2\pi}=0.01$
and $N=150$. The asymptotes for the MSD and time averaged MSD are shown as the dashed lines.}
\label{fig-msd-tamsd-omega}
\end{figure}

\begin{figure}
\includegraphics[width=8.8cm]{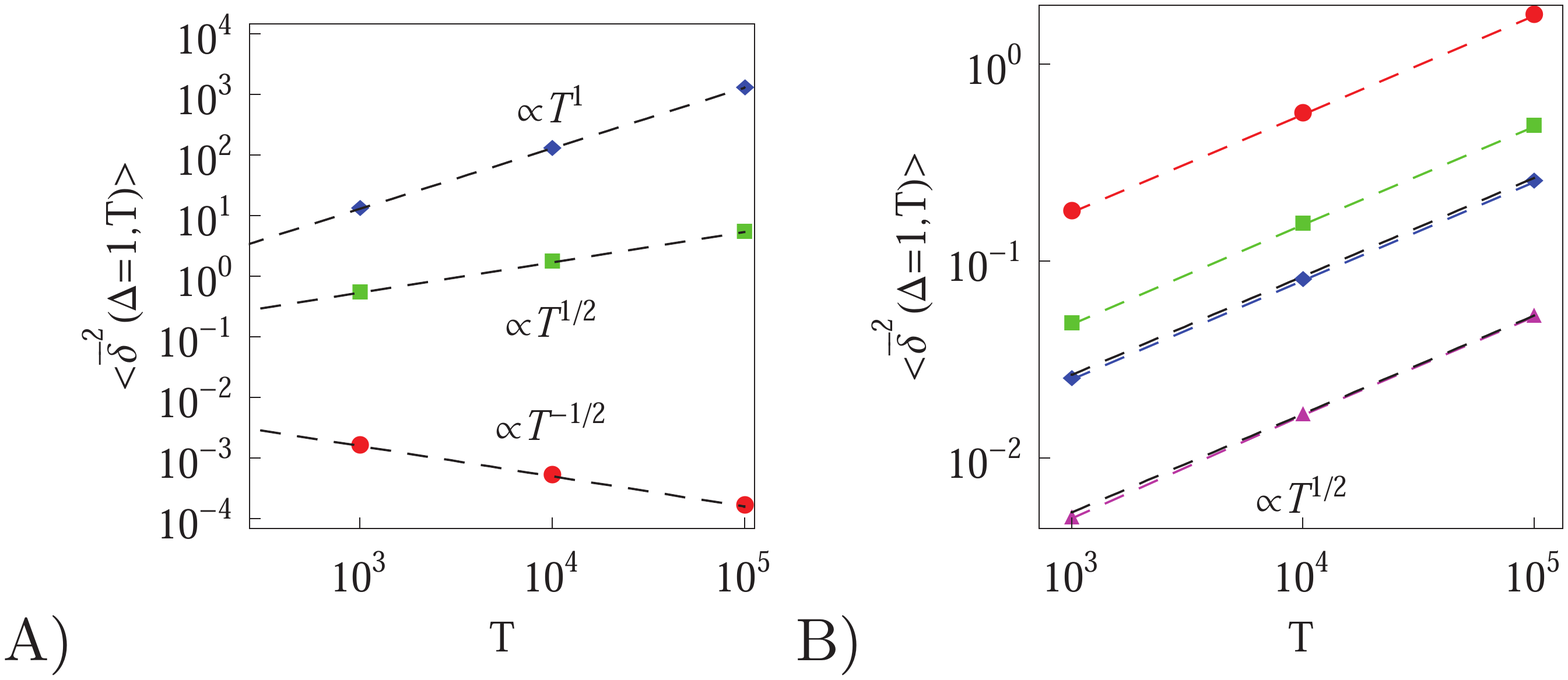}
\caption{Time averaged MSD versus the trace length $T$ for (A)
non-confined overdamped RDPs for $\omega=-1/2,1/2,1$ (for
data points from bottom to top) and for (B) non-confined underdamped RDPs for
$m=1,5,10,50$ (for data points from top to bottom) and $\omega=1/2$. Other
parameters are the same as in Fig. \ref{fig-msd-tamsd} and $2\gamma\mathscr{D}
/\sqrt{2\pi}=0.01$. Dashed lines indicate the scaling relation (\ref{eq-t-dependence-ala-hdps})
in panel A and Eq.~(\ref{eq-initial-ballistic-tamsd-large-beta}) in panel B.}
\label{fig-tamsd-scaling-t}
\end{figure}

\begin{figure*}
\includegraphics[width=18cm]{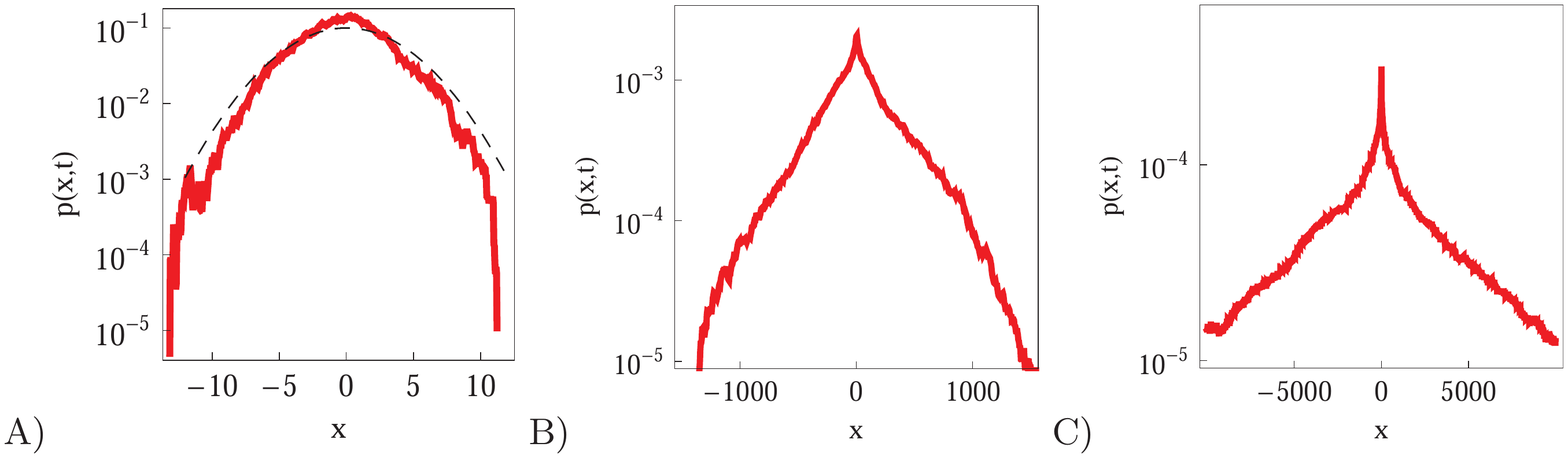}
\caption{PDF of the overdamped RDPs for $\omega=-1/2$, $1/2$, and $1$ (from left
to right). Other parameters are the same as in Fig.~\ref{fig-msd-tamsd} with
$2\gamma\mathscr{D}/\sqrt{2\pi}=0.01$ and $N=300$. The PDF asymptote for 
subdiffusive SBM (\ref{eq-pdf-ala-sbm}) is shown as the dashed curve.}
\label{fig-pdf-spiky}
\end{figure*}

\begin{figure*}
\includegraphics[width=18cm]{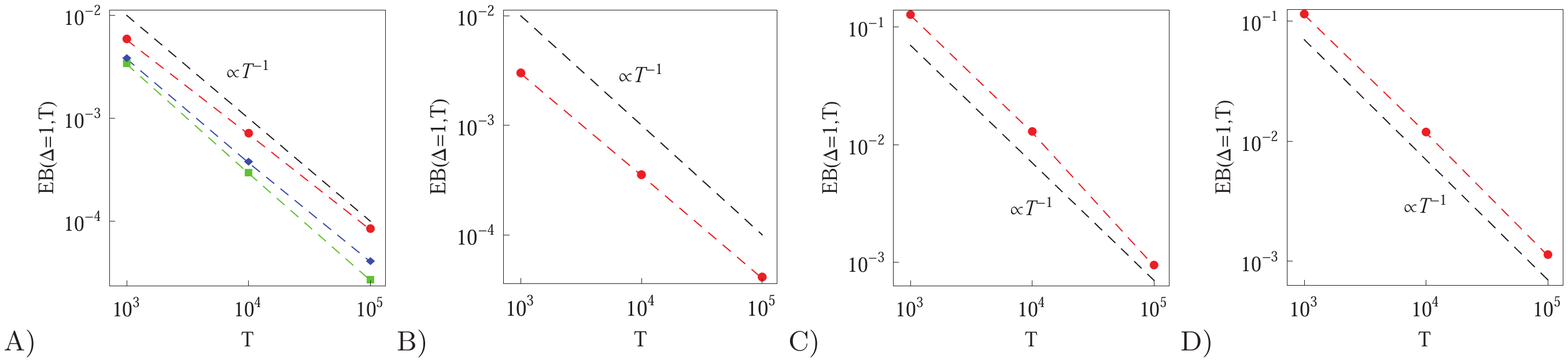}
\caption{EB parameter at $\Delta=1$ computed for free and confined overdamped RDPs
(panels A, B) and for free and confined underdamped RDPs (panels C, D).
Parameters: for all the panels $\omega=1/2$, $2\gamma\mathscr{D}/\sqrt{2\pi}=
0.01$ and (panel A: free overdamped process) $m=0$, $L=\infty$; 
(panel B: confined overdamped process) $m=0$, $L=10^2$; 
(panel C: free underdamped process) $m=50$, $L=\infty$; 
(panel D: confined underdamped process) $m=50$, $L=10^2$.}
\label{fig-eb-confined-t}
\end{figure*}

\begin{figure}
\includegraphics[width=6.8cm]{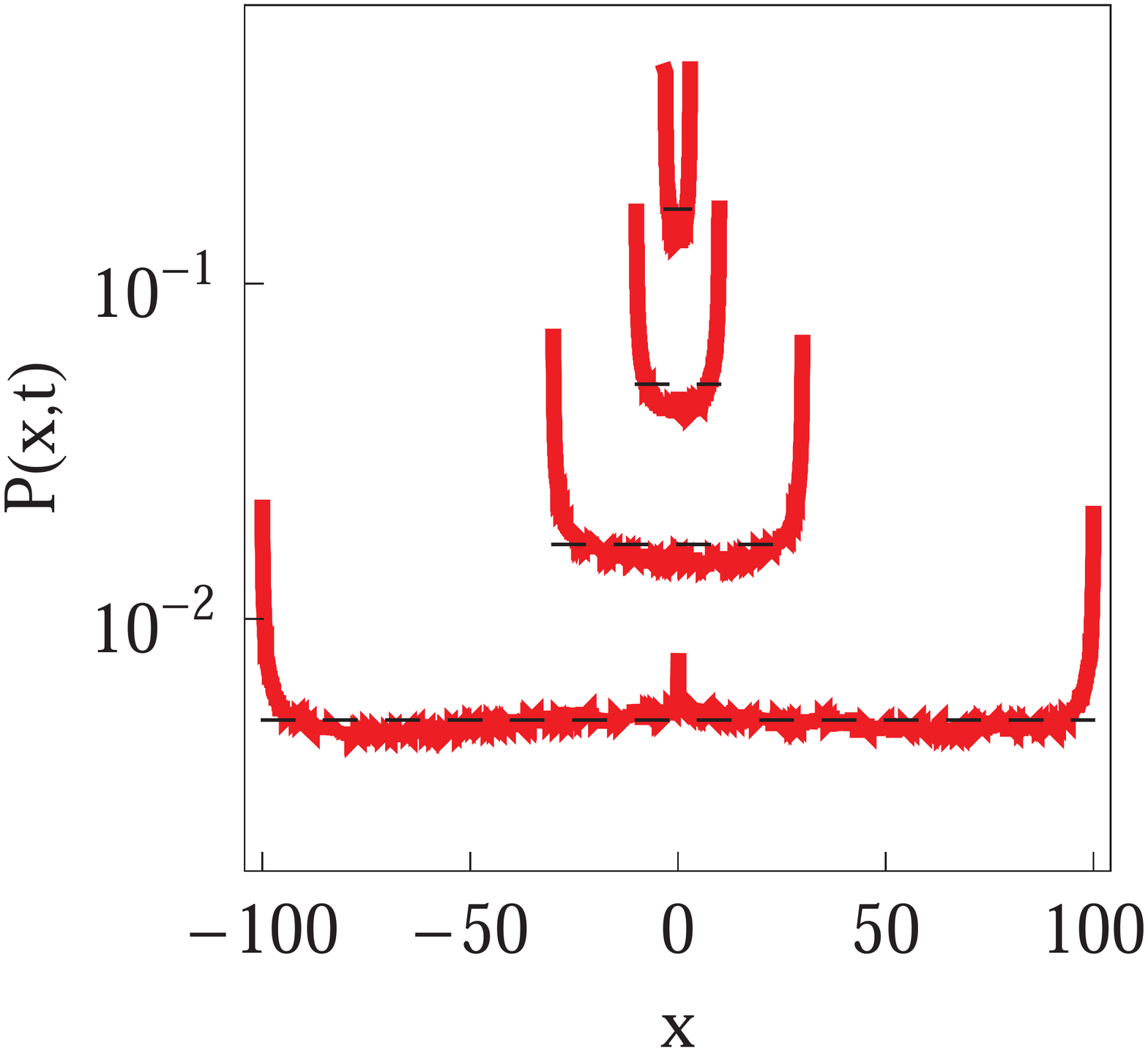}
\caption{PDF of underdamped confined RDPs for $L=3$, $10$, $30$, and $100$ (from
top to bottom), computed for $\omega=1/2$, $2\gamma\mathscr{D}/
\sqrt{2\pi}=0.01$, $T=10^5$, $m=50$. Dashed lines designate the uniform
distribution of particles on the interval.}
\label{fig-pdf-confined-l}
\end{figure}

\begin{figure}
\includegraphics[width=6.8cm]{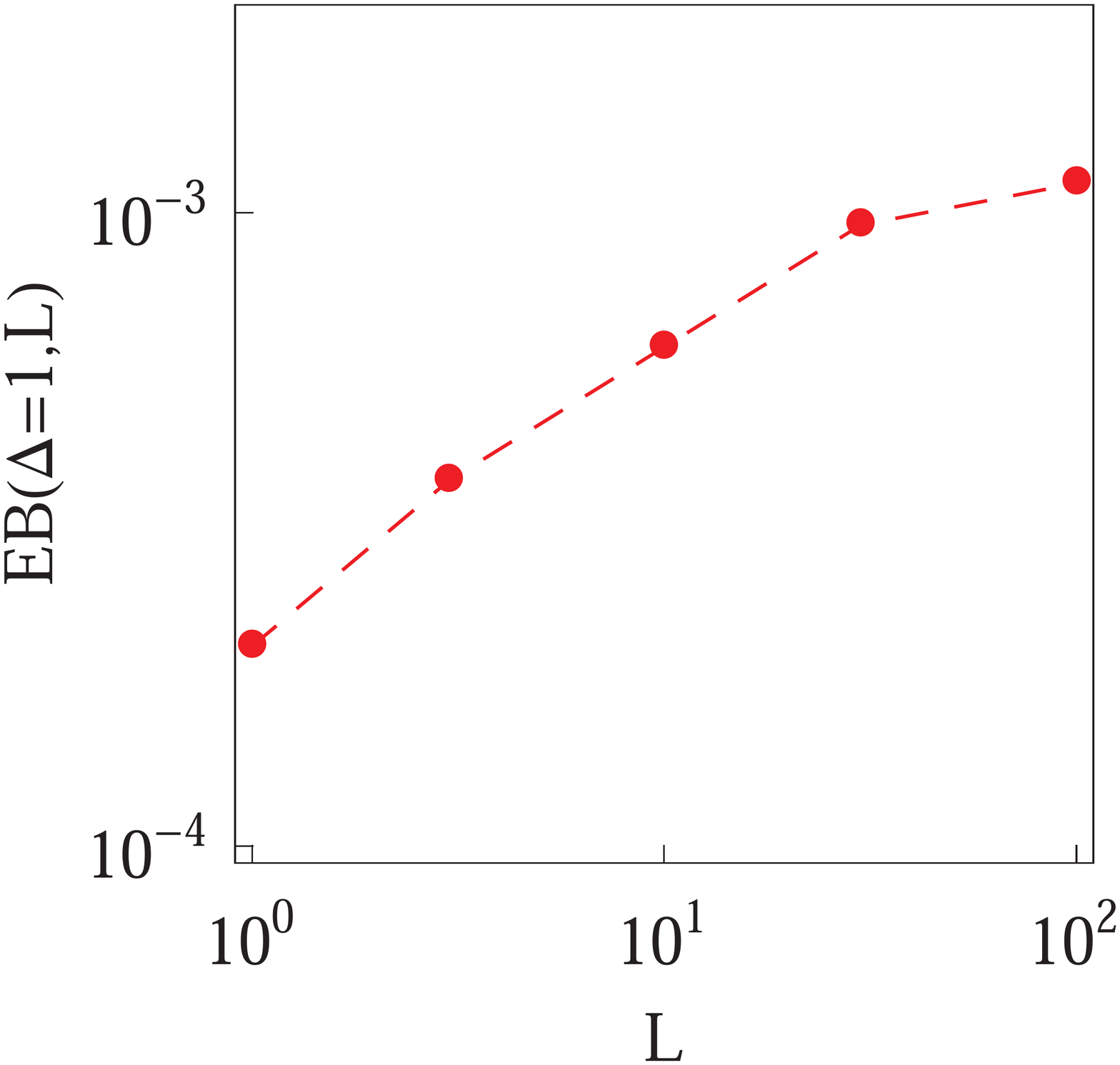}
\caption{EB parameter at $\Delta=1$ for underdamped RDPs with varying
degree of confinement $L$. Parameters are the same as in Fig.
\ref{fig-pdf-confined-l}: $\omega=1/2$, $2\gamma\mathscr{D}/\sqrt{2\pi}=0.01$,
$T=10^5$, $m=50$.}
\label{fig-eb-confined-l}
\end{figure}


\begin{thebibliography}{99}

\bibitem{metz00} J.-P. Bouchaud and A. Georges, Phys. Rep. \textbf{195}, 127
(1990); 
R. Metzler and J. Klafter, Phys. Rep. \textbf{339}, 1 (2000); J. Phys. A \textbf{37}, R161 (2004).

\bibitem{brauchle} C. Br{\"a}uchle, D. C. Lamb, and J. Michaelis, Single Particle
Tracking and Single Molecule Energy Transfer (Wiley-VCH, Weinheim, Germany,
2012); X. S. Xie, P. J. Choi, G.-W. Li, N. K. Lee, and G. Lia, Annu. Rev.
Biophys. \textbf{37}, 417 (2008).

\bibitem{jeney} T. Franosch, M. Grimm, M. Belushkin, F. M. Mor, G. Foffi,
L. Forro, and S. Jeney, Nature \textbf{478}, 7367 (2011).

\bibitem{dense} 
D. S. Banks and C. Fradin, Biophys. J. \textbf{89}, 2960 (2005); 
G. Guigas, C. Kalla and M. Weiss, Biophys. J. \textbf{93}, 316 (2007);
J. Szymanski and M. Weiss, Phys. Rev. Lett. \textbf{103}, 038102 (2009);
W. Pan, L. Filobelo, N. D. Q. Pham, O. Galkin, V. V. Uzunova, and
P. G. Vekilov, Phys. Rev. Lett. \textbf{102}, 058101 (2009);
J.-H. Jeon, N. Leijnse, L. B. Oddershede, and R. Metzler, New
J. Phys. \textbf{15}, 045011 (2013).

\bibitem{metz11} S. Burov, J.-H. Jeon, R. Metzler, and E. Barkai,
Phys. Chem. Chem. Phys. \textbf{13}, 1800 (2011).

\bibitem{metz12} E. Barkai, Y. Garini, and R. Metzler, Phys. Today
\textbf{65(8)}, 29 (2012).

\bibitem{fran13} F. H\"ofling and T. Franosch, Rep. Prog. Phys. \textbf{76},
046602 (2013).

\bibitem{metz14} R. Metzler, J.-H. Jeon, A. G. Cherstvy, and E. Barkai,
Phys. Chem. Chem. Phys. \textbf{16}, 24128 (2014).

\bibitem{soko15} Y. Meroz and I. M. Sokolov, Phys. Rep. \textbf{573}, 1 (2015).

\bibitem{para15} C. Manzo and M. F. Garcia-Parajo, Rep. Prog. Phys. \textbf{78},
124601 (2015).

\bibitem{cell} A. Caspi, R. Granek, and M. Elbaum, Phys. Rev. E \textbf{66},
011916 (2002); 
G. Seisenberger et al., Science \textbf{294}, 1929 (2001);
I. Golding and E. C. Cox, Phys. Rev. Lett. \textbf{96}, 098102 (2006); 
I. Bronstein, Y. Israel, E. Kepten, S. Mai, Y. Shav-Tal, E. Barkai, and Y. Garini,
Phys. Rev. Lett. \textbf{103}, 018102 (2009);
S. M. A. Tabei, S. Burov, H. Y. Kim, A. Kuznetsov, T. Huynh, J. Jureller, L. H.
Philipson, A. R. Dinner, and N. F. Scherer, Proc. Natl. Acad. Sci. U. S. A. \textbf{110}, 4911
(2013).

\bibitem{weis04} M. Weiss, M. Elsner, F. Kartberg, and T. Nilsson, Biophys. J.
\textbf{87}, 3518 (2004).

\bibitem{metz11b} J.-H. Jeon, V. Tejedor, S. Burov, E. Barkai, C. Selhuber-Unkel,
K. Berg-Sorensen, L. Oddershede, and R. Metzler, Phys. Rev. Lett. \textbf{106},
048103 (2011).

\bibitem{cellsuper} A. Caspi, R. Granek, and M. Elbaum, Phys. Rev. Lett.
\textbf{85}, 5655 (2000);
N. Gal and D. Weihs, Phys. Rev. E {\bf 81}, 020903(R) (2010);
D. Robert, T. H. Nguyen, F. Gallet, and C. Wilhelm, PLoS ONE
\textbf{4}, e10046 (2010);
J. F. Reverey, J.-H. Jeon, H. Bao, M. Leippe, R. Metzler, and
C. Selhuber-Unkel, Sci. Rep. \textbf{5}, 11690 (2015).

\bibitem{lipids} G. R. Kneller, K. Baczynski, and M. Pasienkewicz-Gierula,
J. Chem. Phys. \textbf{135}, 141105 (2011);
T. Akimoto, E. Yamamoto, K. Yasuoka, Y. Hirano, and M. Yasui, Phys. Rev. Lett.
\textbf{107}, 178103 (2011);
J.-H. Jeon, H. Martinez-Seara Monne, M. Javanainen, and R. Metzler, Phys. Rev.
Lett. \textbf{109}, 188103 (2012);
M. Javanainen, H. Hammaren, L. Monticelli, J.-H. Jeon, R. Metzler, and I.
Vattulainen, Faraday Discussions \textbf{161}, 397 (2013);
S. Stachura and G. R. Kneller, J. Chem. Phys. \textbf{40}, 245 (2014);
E. Yamamoto, A. C. Kalli, T. Akimoto, K. Yasuoka, and M. S. P. Sansom, Sci. Rep.
\textbf{5}, 18245 (2015).

\bibitem{prx} J.-H. Jeon, M. Javanainen, H. Martinez-Seara, R. Metzler, and I.
Vattulainen, Phys. Rev. X \textbf{6}, 021006 (2016).

\bibitem{membranes} A. V. Weigel, B. Simon, M. M. Tamkun and D. Krapf, Proc.
Natl. Acad. Sci. U. S. A. \textbf{108}, 6438 (2011);
D. Krapf, Curr. Topics Membr. \textbf{75}, 167 (2015); 
D. Krapf, G. Campagnola, K. Nepal, and O. B. Peersen, Phys. Chem. Chem. Phys. \textbf{18}, 12633 (2016).

\bibitem{metz16} R. Metzler, J.-H. Jeon, and A. G. Cherstvy, Biochem. Biophys. (2016). 
Acta doi:10.1016/j.bbamem.2016.01.022.

\bibitem{lape15} C. Manzo, J. A. Torreno-Pina, P. Massignan, G. J. Lapeyre, Jr.,
M. Lewenstein, and M. F. Garcia Parajo, Phys. Rev X \textbf{5}, 011021 (2015).

\bibitem{crowd} H. Berry and H. A. Soula, Front. Physiol. \textbf{5}, 437 (2014); 
H. Berry and H. Chate, Phys. Rev. E \textbf{89}, 022708 (2014); 
F. Trovato and V. Tozzini, Biophys. J. \textbf{107}, 2579 (2014);
M. Weiss, Intl. Rev. Cell \& Molec. Biol. \textbf{307}, 383 (2014); 
M. J. Saxton, J. Phys. Chem. B \textbf{118}, 12805 (2014);
D. S. Banks, C. Tressler, R. D. Peters, F. H\"ofling, and C. Fradin,
Soft Matter \textbf{12}, 4190 (2016);
T. Sentjabrskaja,  E. Zaccarelli, C. De Michele, F. Sciortino, P. Tartaglia,
F. H{\"o}fling and T. Franosch, Phys. Rev. Lett. \textbf{98},
140601 (2007);
S. Leitmann and T. Franosch, Phys. Rev. Lett. \textbf{111},
190603 (2013);
T. Voigtmann, S. U. Egelhaaf, and M. Laurati, Nature Comm. \textbf{7}, 11133
(2016).

\bibitem{ghos15} S. K. Ghosh, A. G. Cherstvy, and R. Metzler, Phys. Chem. Chem. Phys.
\textbf{17}, 472 (2015).

\bibitem{ghos16} S. Ghosh, A. G. Cherstvy, D. S. Grebenkov, and R. Metzler, New. J. Phys.
\textbf{16}, 013027 (2016).


\bibitem{rogers} S. S. Rogers, C. van der Walle, and T. A. Weigh, Langmuir
\textbf{24}, 13549 (2008).

\bibitem{hydra} A. Upadhyaya, J.-P. Rieu, J. A. Glazier, and Y. Sawada, Physica
A \textbf{293}, 549 (2001);
L. G. A. Alves, D. B. Scariot, R. R. Guimaraes, C. V. Nakamura, R. S. Mendes,
and H. V. Ribeiro, PLoS One \textbf{11}, e0152092 (2016).

\bibitem{mung16} I. Munguira, I. Casuso, H. Takahashi, F. Rico, A. Miyagi, M.
Chami, and S. Scheuring, ACS Nano \textbf{10}, 2584 (2016).

\bibitem{moss16} S. Hanot, S. Lyonnard, and S. Mossa, Nanoscale \textbf{8}, 3314
(2016).

\bibitem{schw15} D. Wang, C. He, M. P. Stoykovich, and D. K. Schwartz, ACS Nano
\textbf{9}, 1656 (2015).

\bibitem{schw13} M. J. Skaug, J. Mabry, and D. K. Schwartz, Phys. Rev. Lett.
\textbf{110}, 256101 (2013).

\bibitem{yeth14} G. Kwon, B. J. Sung, and A. Yethiraj, J. Phys. Chem. B
\textbf{118}, 8128 (2014).

\bibitem{week00} E. R. Weeks, J. C. Crocker, A. C. Levitt, A. Schofield, and D. A.
Weitz, Science \textbf{287}, 627 (2000).

\bibitem{zamp14} P. Charbonneau, Y. Jin, G. Parisi, and F. Zamponi, Proc. Natl.
Acad. Sci. U. S. A. \textbf{111}, 15025 (2014).

\bibitem{schw15b} M. J. Skaug, L. Wang, Y. Ding, and D. K. Schwartz, ACS Nano
\textbf{9}, 2148 (2015); 
M. J. Skaug and D. K. Schwartz, Ind. Eng. Chem. Res. \textbf{54}, 4414 (2015).

\bibitem{berk06} B. Berkowitz, A. Cortis, M. Dentz, and H. Scher, 
Rev. Geophys. \textbf{44}, RG2003 (2006). 

\bibitem{gran09} B. Wang, S. M. Anthony, S. C. Bae, and S. Granick, Proc. Natl.
Acad. Sci. U.S.A. \textbf{106}, 15160 (2009).

\bibitem{gran12} B. Wang, J. Kuo, S. C. Bae, and S. Granick, Nature Mater.
\textbf{11}, 481 (2012).

\bibitem{gran14} J. Guan, B. Wang, and S. Granick, ACS Nano \textbf{8}, 3331 (2014).

\bibitem{kris13} K. He, F. B. Khorasani, S. T. Retterer, D. K. Thomas, J. C.
Conrad, and R. Krishnamoorti, ACS Nano \textbf{7}, 5122 (2013). 

\bibitem{mont65} 
E. W. Montroll and G. H. Weiss, J. Math. Phys. \textbf{6}, 167
(1965); 
H. Scher and E. W. Montroll, Phys. Rev. B \textbf{12}, 2455 (1975).

\bibitem{noisy} J.-H. Jeon, E. Barkai, and R. Metzler, J. Chem. Phys. \textbf{139},
121916 (2013).

\bibitem{traps} T. Akimoto and E. Barkai, Phys. Rev. E \textbf{87}, 032915 (2013);
T. Akimoto, Phys. Rev. Lett. \textbf{108}, 164101 (2012);
T. Geisel and S. Thomae, Phys. Rev. Lett. \textbf{52}, 1936 (1984);
T. Geisel, J. Nierwetberg and A. Zacherl, Phys. Rev. Lett. \textbf{54}, 616 (1985);
M. W. Deem and D. Chandler, J. Stat. Phys. \textbf{76}, 911 (1994);
C. Monthus and J.-P. Bouchaud, J. Phys. A: Math. Gen. \textbf{29}, 3847 (1996);
E. Bertin and J.-P. Bouchaud, Phys. Rev. E \textbf{67}, 026128 (2003);
S. Burov and E. Barkai, Phys. Rev. Lett. \textbf{98}, 250601 (2007);
M. Dentz et. al., Adv. Water Res. \textbf{49}, 13 (2012);
T. Akimoto, E. Barkai, and K. Saito, arXiv:1604.06175.

\bibitem{corr} M. M. Meerschaert, E. Nane, and Y. Xiao, Stat. Probab. Lett.
\textbf{79}, 1194 (2009);
A. V. Chechkin, M. Hofmann, and I. M. Sokolov, Phys. Rev. E \textbf{80}, 031112
(2009);
J. H. P. Schulz, A. V. Chechkin and R. Metzler, J. Phys. A. \textbf{46}, 475001
(2013).

\bibitem{corr1} V. Tejedor and R. Metzler, J. Phys. A \textbf{43}, 082002 (2010);
M. Magdziarz, R. Metzler, W. Szczotka, and P. Zebrowski, Phys. Rev. E \textbf{85},
051103 (2012);

\bibitem{fbm} B. B. Mandelbrot and J. W. van Ness, SIAM Rev. \textbf{10},
422 (1968).

\bibitem{fle} R. Kubo, Rep. Prog. Phys. \textbf{29}, 255 (1966);
P. H{\"a}nggi, Zeit. Physik B \textbf{31}, 407 (1978); 
P. H{\"a}nggi and F. Mojtabai, Phys. Rev. E \textbf{26}, 1168 (1982); 
S. C. Kou, Ann. Appl. Stat. \textbf{2}, 501 (2008);
I. Goychuk, Phys. Rev. E \textbf{80}, 046125 (2009); 
I. Goychuk, Adv. Chem. Phys. \textbf{150}, 187 (2012).

\bibitem{pagn16} G. Pagnini, D. Molina-Garcia, T. M. Pham, C. Manzo, and P.
Paradisi, E-print arXiv:1508.01361.

\bibitem{bark09} W. Deng and E. Barkai, Phys. Rev. E \textbf{79}, 011112 (2009).

\bibitem{kehr87} J.W. Haus and K. W. Kehr, Phys. Rep. \textbf{150}, 263 (1987);
S. Havlin and D. Ben--Avraham, Adv. Phys. {\bf 36}, 695 (1987);
G. C. Papanicolaou, "Diffusion in random media", Surveys in applied mathematics,
pp. 205-253 (New York, Plenum Press, 1995);
G. M. Zaslavsky, Phys. Rep. \textbf{371}, 461 (2002); 
J. P. Bouchaud, A. Comtet, A. Georges, and P. le Doussal, Annal. Phys.
\textbf{201}, 285 (1990).

\bibitem{bray13} A. J. Bray, S. N. Majumdar, G. Schehr, Adv. Phys. \textbf{62}, 225 (2013).

\bibitem{mard15} Y. Mardoukhi, J.-H. Jeon, and R. Metzler, Phys Chem Chem Phys.
\textbf{17}, 30134 (2015).

\bibitem{lape14} P. Massignan et al., Phys. Rev. Lett. \textbf{112}, 150603 (2014).

\bibitem{slat14} M. V. Chubynsky and G. W. Slater, Phys. Rev. Lett. \textbf{113},
098302 (2014).

\bibitem{akim15} T. Uneyama, T. Miyaguchi, and T. Akimoto, Phys. Rev. E
\textbf{92}, 032140 (2015).

\bibitem{seba16} R. Jain and K. L. Sebastian, J. Phys. Chem. B \textbf{120}, 3988 (2016).

\bibitem{egel16} J. Bewerunge, I. Ladadwa, F. Platten, C. Zunke, A. Heuer, and S. Egelhaaf, Phys. Chem. Chem. Phys. \textbf{18}, 18887 (2016).

\bibitem{gotz92} W. Gotze and L. Sjogren, Rep. Prog. Phys. \textbf{55}, 241 (1992).

\bibitem{yama98} R. Yamamoto and A. Onuki, Phys. Rev. Lett. \textbf{81}, 4915
(1998).

\bibitem{rich02} R. Richert, J. Phys.: Condens. Matter \textbf{14}, R703 (2002).

\bibitem{biro11} L. Berthier and G. Biroli, Rev. Mod. Phys. \textbf{83}, 587 (2011).

\bibitem{saxt97} M. J. Saxton, Biophys. J. \textbf{72}, 1744 (1997).

\bibitem{platani} M. Platani, I. Goldberg, A. I. Lamond, and J. R. Swedlow,
Nat. Cell Biol. \textbf{4}, 502 (2002).

\bibitem{aust06} Y. M. Wang, R. H. Austin, and E. C. Cox, Phys. Rev. Lett. \textbf{97}, 048302 (2006).

\bibitem{goyc14} I. Goychuk and V. O. Kharchenko, Phys. Rev. Lett. \textbf{113},
100601 (2014).

\bibitem{baue15} M. Bauer, E. S. Rasmussen, M. A. Lomholt, and R. Metzler, Sci.
Rep. \textbf{5}, 10072 (2015).

\bibitem{hapc09} S. Hapca, J. W. Crawford, and L. M. Young, J. R. Soc. Interface
\textbf{6}, 111 (2009).

\bibitem{yama02} K. Yamamura, Popul. Ecol. \textbf{44}, 93 (2002).

\bibitem{clar98} J. S. Clark, Amer. Naturalist \textbf{152}, 204 (1998).

\bibitem{nath06} R. Nathan, Science \textbf{313}, 786 (2006).

\bibitem{yama04} K. Yamamura, Popul. Ecol. \textbf{46}, 87 (2004);

\bibitem{skal00} G. T. Skalski and J. F. Gilliam, Ecology \textbf{81}, 1685 (2000).

\bibitem{hovm02} J. K. M. Brown and M. S. Hovmoller, Science Translat. Medic.
\textbf{297}, 537 (2002).

\bibitem{lang11} T. K\"uhn, T. O. Ihalainen, J. Hyvaluoma, N. Dross, S. F. Willman, 
J. Langowski, M. Vihinen-Ranta, and J. Timonen, PLoS One \textbf{6}, e22962 (2011).

\bibitem{elf11} B. P. English, V. Hauryliuk, A. Sanamrad, S. Tankov, N. H. Dekker,
and J. Elf, Proc. Natl. Acad. Sci. U. S. A. \textbf{108}, E365 (2011).

\bibitem{brau13} C. B. Mast, S. Schink, U. Gerland, and D. Braun, Proc. Natl. Acad.
Sci. U. S. A. \textbf{110}, 8030 (2013).

\bibitem{cher13} A. G. Cherstvy, A. V. Chechkin, and R. Metzler, New J. Phys.
\textbf{15}, 083039 (2013); 
A. G. Cherstvy and R. Metzler, Phys. Chem. Chem. Phys. \textbf{15}, 20220 (2013).

\bibitem{cher14a} A. G. Cherstvy, A. V. Chechkin, and R. Metzler, Soft Matter
\textbf{10}, 1591 (2014).

\bibitem{cher14b} A. G. Cherstvy, A. V. Chechkin, and R. Metzler, J. Phys. A
\textbf{47}, 485002 (2014).

\bibitem{lenz03} K. S. Fa and E. K. Lenzi, Phys. Rev. E \textbf{67}, 061105 (2003). 

\bibitem{fa05} K. S. Fa, Phys. Rev. E \textbf{72}, 020101(R) (2005).

\bibitem{zimm11} M. Burgis et al., New J. Phys. \textbf{13}, 043031 (2011).

\bibitem{heid14} M. Heidern\"atsch, PhD Thesis, "On the diffusion in inhomogeneous systems", TU Chemnitz (2015).

\bibitem{safd15} H. Safdari, A. G. Cherstvy, A. V. Chechkin, F. Thiel, I. M.
Sokolov, and R. Metzler, J. Phys. A \textbf{48}, 375002 (2015).

\bibitem{grig16} E. Geneston, R. Tuladhar, M. T. Beig, M. Bologna, and P.
Grigolini, arXiv:1601.02879.

\bibitem{muni02} S. C. Lim and S. V. Muniandy, Phys. Rev. E \textbf{66}, 021114
(2002).

\bibitem{pein11} R. Friedrich, J. Peinke, M. Sahimi, and M. R. R. Tabar, Phys.
Rep. \textbf{506}, 87 (2011).

\bibitem{bodr15gg} A. Bodrova, A. V. Chechkin, A. G. Cherstvy, and R. Metzler, 
Phys. Chem. Chem. Phys. \textbf{17}, 21791 (2015).

\bibitem{soko14} F. Thiel and I. M. Sokolov, Phys. Rev. E \textbf{89}, 012115
(2014).

\bibitem{jeon14} J.-H. Jeon, A. V. Chechkin, and R. Metzler, Phys. Chem. Chem.
Phys. \textbf{16}, 15811 (2014).

\bibitem{batc52} G. K. Batchelor, Math. Proc. Cambridge Philos. Soc. \textbf{48},
345 (1952).

\bibitem{cher15c} A. G. Cherstvy and R. Metzler, J. Chem. Phys. \textbf{142},
144105 (2015).

\bibitem{fuli13} A. Fulinski, J. Chem. Phys. \textbf{138}, 021101 (2013); Acta Phys. Polon. \textbf{44}, 1137 (2013).

\bibitem{fuli11} A. Fulinski, Phys. Rev. E \textbf{83}, 061140 (2011).

\bibitem{cher15b} A. G. Cherstvy and R. Metzler, J. Stat. Mech. P05010 (2015).

\bibitem{jeon15} H. Safdari, A. V. Chechkin, G. R. Jafari, and R. Metzler, Phys. 
Rev. E \textbf{91}, 042107 (2015).

\bibitem{cher14c}  A. G. Cherstvy and R. Metzler, Phys. Rev. E \textbf{90},
012134 (2014).

\bibitem{lube07} A. W. C. Lau and T. C. Lubensky, Phys. Rev. E \textbf{76},
011123 (2007).

\bibitem{bodr15} A. Bodrova, A. V. Chechkin, A. G. Cherstvy, and R. Metzler,
New J. Phys. \textbf{17}, 063038 (2015).

\bibitem{uhle30} G. E. Uhlenbeck and L. S. Ornstein, Phys. Rev. \textbf{36}, 823
(1930).

\bibitem{bodr16} A. Bodrova, A. V. Chechkin, A. G. Cherstvy, H. Safdari, I. M.
Sokolov, and R. Metzler, Sci. Rep. \textbf{???}, ??? (2016). 

\bibitem{soko16} B. Lindner and I. M. Sokolov, Phys. Rev. E \textbf{93}, 042106 (2016).

\bibitem{rien14} C. Di Rienzo, V. Piazza, E. Gratton, F. Beltram, and F. Cardarelli,
Nature Comm. \textbf{5}, 5891 (2014).

\bibitem{fara17} S. Regev, N. Gronbech-Jensen, and O. Farago, arXiv:1606.08632.

\bibitem{fara14} O. Farago and N. Gronbech-Jensen, Phys. Rev. E \textbf{89},
013301 (2014).

\bibitem{fara14b} O. Farago and N. Gronbech-Jensen, J. Stat. Phys. \textbf{156},
1093 (2014).

\bibitem{bouchaud_web} J.-P. Bouchaud, J. Phys. (Paris) I \textbf{2}, 1705 (1992).

\bibitem{weak} G. Bel and E. Barkai, Phys. Rev. Lett. \textbf{94}, 240602 (2005); 
A. Rebenshtok, and E. Barkai, Phys. Rev. Lett. \textbf{99}, 210601 (2007).

\bibitem{soko08} A. Lubelski, I. M. Sokolov, and J. Klafter, Phys. Rev. Lett.
\textbf{100}, 250602 (2008).

\bibitem{buro10} S. Burov, R. Metzler, and E. Barkai, Proc. Natl. Acad. Sci.
U. S. A. \textbf{107}, 13228 (2010).

\bibitem{metz08} Y. He, S. Burov, R. Metzler, and E. Barkai, Phys. Rev. Lett.
\textbf{101}, 058101 (2008).

\bibitem{jeon10} J.-H. Jeon and R. Metzler, Phys. Rev. E \textbf{81}, 021103 (2010).

\bibitem{jeon13} J.-H. Jeon, N. Leijnse, L. B. Oddershede, and R. Metzler, New J.
Phys. \textbf{15}, 045011 (2013).

\bibitem{gode16} M. Schwarzl, A. Godec, E. Barkai, and R. Metzler (unpublished).

\bibitem{greb12} A. Andreanov and D. S. Grebenkov, J. Stat. Mech. P07001 (2012).

\bibitem{kurs13} J. Kursawe, J. Schulz, and R. Metzler, Phys. Rev. E \textbf{88},
062124 (2013).

\bibitem{gode13}  A. Godec and R. Metzler, Phys. Rev. Lett. \textbf{110}, 020603
(2013).

\bibitem{metz10} J.-H. Jeon and R. Metzler, J. Phys. A \textbf{43}, 252001 (2010).

\bibitem{goyc10} P. Siegle, I. Goychuk, and P. H\"anggi, Phys. Rev. Lett.
\textbf{105}, 100602 (2010).

\bibitem{lind07} B. Lindner, New J. Phys. \textbf{9}, 136 (2007); J. Stat. Phys. \textbf{130}, 523 (2008); New J. Phys. \textbf{12}, 063026 (2010).

\bibitem{roma12} P. Romanczuk, M. B\"ar, W. Ebeling, B. Lindner, and L. Schimansky-Geier, Eur. Phys. J. Special Topics \textbf{202}, 1 (2012).

\bibitem{saxt01} M. J. Saxton, Biophys. J. \textbf{81}, 2226 (2001).

\bibitem{webb96} T. J. Feder, I. Brust-Mascher, J. P. Slattery, B. Baird, and W. W. Webb, Biophys. J. \textbf{70}, 2767 (1996).

\bibitem{webb99} P. Schwille, J. Korlach, and W. W. Webb, Cytometry \textbf{36}, 176 (1999).

\bibitem{weis03} M. Weiss, H. Hashimoto, and T. Nilsson, Biophys. J. \textbf{84}, 4043 (2003).

\bibitem{kusu93} A. Kusumi, Y. Sako, and M. Yamamoto, Biophys. J. \textbf{65}, 2021 (1993).


\bibitem{hapc07} S. Hapca, J. W. Crawford, K. MacMillan, M. J. Wilson, and I. M. Young, J. Theor. Biol. \textbf{248}, 212 (2007).

\bibitem{para16} C. Charalambous, G. Munoz-Gil, A. Celi, M. F. Garcia-Parajo, M. Lewenstein, C. Manzo, and M. A. Garcıa-March, arXiv:1607.00189.

\bibitem{cher17} A. G. Cherstvy, A. V. Chechkin, and R. Metzler, unpublished.

\end{thebibliography}
\end{document}